\documentclass[final,1p,times]{elsarticle}
\usepackage{tikz}

\usepackage{comment}
\usepackage{amsmath}
\usepackage{amsfonts}
\usepackage{multirow}
\usepackage{graphicx}
\usepackage[colorlinks=true, allcolors=blue]{hyperref}
\usepackage{booktabs}
\usepackage{longtable}
\usepackage{geometry}
\usepackage{lscape}
\usepackage{pdflscape}
\usepackage{caption}
\usepackage{subcaption}
\usepackage{array}
\usepackage{geometry}
\usepackage{bm}
\usepackage{float}
\usepackage{microtype}
\usepackage{hyperref}

\journal{abc}
\begin{document}

\begin{frontmatter}

\title{Shrinkage Estimators for Mean and Covariance: Evidence on Portfolio Efficiency Across Market Dimensions}

\author{Rupendra Yadav$^{a}$, Amita Sharma$^{b}$, Aparna Mehra$^{a}$}

 \address[${a}$]{Department of Mathematics, Indian Institute of Technology Delhi, Hauz Khas, New Delhi - 110016, India}
\address[$b$]{Department of  Mathematics, Netaji Subhash University of Technology, Dwarka Sec 3, New Delhi -110078, India }
\begin{abstract}
The mean-variance model remains the most prevalent investment framework, built on diversification principles. However, it consistently struggles with estimation errors in expected returns and the covariance matrix, its core parameters. To address this concern, this research evaluates the performance of mean-variance (MV) and global minimum-variance (GMV) models across various shrinkage estimators designed to improve these parameters. Specifically, we examine five shrinkage estimators for expected returns and eleven for the covariance matrix. 
To compare multiple portfolios, we employ a super-efficient data envelopment analysis model to rank the portfolios according to investors' risk-return preferences.  Our comprehensive empirical investigation utilizes six real-world datasets with different dimensional characteristics, applying a rolling window methodology across three out-of-sample testing periods.  Following the ranking process, we examine the chosen shrinkage-based MV or GMV portfolios against five traditional portfolio optimization techniques—classical MV and GMV for sample estimates, MiniMax, conditional value-at-risk, and semi-mean absolute deviation risk measures. Our empirical findings reveal that, in most scenarios, the GMV model combined with the Ledoit-Wolf two-parameter shrinkage covariance estimator (COV2) represents the optimal selection for a broad spectrum of investors.  Meanwhile, the MV model utilizing COV2 alongside the sample mean (SM) proves more suitable for return-oriented investors. These two identified models demonstrate superior performance compared to traditional benchmark approaches. Overall, this study lays the groundwork for a more comprehensive understanding of how specific shrinkage models perform across diverse investor profiles and market setups.

\end{abstract}

\end{frontmatter}

\textbf{Highlights}
\begin{itemize} 
\item Deliver empirical evidence on the practical value and benefits of shrinkage estimators in portfolio selection.

\item Integrates shrinkage estimators into mean-variance and global minimum-variance optimization frameworks.

\item Evaluate the effectiveness of shrinkage estimators across diverse low- and high-dimensional emerging and developed financial markets.

\item Employing a super-efficiency DEA model to rank the shrinkage-based optimal portfolios. 

\item Observed that the two-parameter shrinkage covariance estimator outperforms other benchmark models.
\end{itemize}
\section{Introduction}
\subsection{Motivation and Background}

Since its introduction, modern portfolio theory (MPT) \cite{markowitz1952portfolio} has been a cornerstone of quantitative finance, providing a formal framework for creating portfolios that balance the trade-off between mean return and risk. A core of MPT is diversification, which aims to reduce overall portfolio risk. 
However, recent global financial stresses have highlighted flaws in conventional portfolio models, especially those that rely on assumptions of stable return distributions and fail to account for severe market events. During periods of high uncertainty, the relationship between return and risk can become distorted, challenging the reliability of the standard Markowitz mean-variance framework.

A persistent challenge in implementing Mean-Variance Optimization (MVO) is the difficulty in accurately estimating mean returns and the covariance matrix of asset returns \cite{chopra1993effect}. Errors in these estimations can significantly distort portfolio weights, often leading to illogical or overly concentrated allocations. Furthermore, in practice, institutional or regulatory rules often prohibit short selling, requiring new models that can incorporate these limitations while still ensuring adequate diversification. 

The mean-variance optimization (MVO) framework has several documented shortcomings. MVO portfolios frequently deliver worse performance than simpler strategies like equal weighting \cite{demiguel2009optimal, jobson1981putting} and often demonstrate poor diversification \cite{Green1992}. A core issue, as noted by Michaud \cite{michaud2001efficient}, is that the mean-variance optimizer is overly sensitive to inaccuracies in the input estimates, resulting in unstable and difficult-to-interpret results.
Furthermore, MVO typically relies on deterministic point estimates for means and covariances. This methodology fails to account for the inherent uncertainty in estimating these parameters, which can consequently produce unreliable asset allocations.

The global minimum variance (GMV) model, which ignores mean estimation errors, is one approach to reducing estimation errors in the mean-variance (MV) model. This is because it is generally accepted that estimation error in the mean is more significant than that resulting from covariance terms \cite{chopra1993effect,best2015sensitivity}. Even though GMV approach is straightforward, it still faces issues, particularly when return distributions have heavy tails, which can lead to significant estimation errors in the covariance as well \cite{martin2009asset}. 

Alternatively, researchers impose a weight norm constraint \cite{demiguel2009optimal} in the MV model or explore translating it into its robust version by defining uncertainty sets for the mean vector and covariance matrix \cite{lee2020sparse}. While robust optimization provides a formal way to address input uncertainty, it is not without criticism.  Scherer \cite{scherer2007can} argues that robust framework often behaves similarly to applying shrinkage to input estimates and, in some cases, may even result in poorer out-of-sample performance. This critique has reinforced interest in directly improving estimation techniques, particularly in high-dimensional settings where classical estimators become unstable due to the `curse of dimensionality' \cite{bai2010spectral, bodnar2019testing}. 

Shrinkage estimators have shown particular promise in addressing this issue. By blending sample estimates with structured priors, they produce more reliable and stable inputs for the MV model, especially when the number of assets exceeds the number of observations. This stability is crucial for high-dimensional portfolios, for example, the S\&P 500 or the RUSSELL 1000 market indices, where minor estimation errors can lead to significant performance deterioration. 

Several shrinkage estimators for the population mean and population covariance matrix are available in the literature. For the population mean, notable contributions include the James-Stein shrinkage estimator (JS) \cite{james1961estimation} that improves multivariate mean estimation by shrinking sample estimates toward a common value, excelling in high-dimensional settings. Building on this, Jorion's Bayes-Stein shrinkage estimator (BS) \cite{jorion1986bayes} incorporates a Bayesian framework, adapting shrinkage based on prior information, which enhances mean estimation, particularly in financial contexts. The quadratic shrinkage estimator (QUAD) \cite{wang2014non} further refines the shrinkage process by applying a quadratic shrinkage function. Following this, Bodnar et al. \cite{bodnar2019optimal} introduced the Bodnar optimal linear shrinkage estimator (BOP), which optimizes linear shrinkage to minimize the mean squared error, offering asymptotically stable and accurate results in true mean estimation. 

In parallel, the literature on shrinkage estimators for covariance matrices was advanced by Ledoit and Wolf \cite{ledoit2003improved,ledoit2004well} for the development of linear shrinkage estimators for large covariance matrices, introducing a computationally efficient and robust approach that improves estimation accuracy in high-dimensional settings by shrinking the sample covariance matrix toward a structured target. The work of Bodnar et al. \cite{bodnar2014strong} presents the construction of an optimal linear shrinkage estimator (OLSE) for the covariance matrix, which is shown to asymptotically minimize the Frobenius loss with almost sure convergence.

Building on their earlier work, Ledoit and Wolf \cite{ledoit2012nonlinear, ledoit2015spectrum} introduced a nonlinear shrinkage estimator using an indirect approach that recovers the population eigenvalues of the covariance matrix, achieving more precise and adaptive shrinkage, particularly when the sample size is small relative to the number of assets.
In a further advancement, Ledoit and Wolf \cite{ledoit2020analytical} proposed a nonlinear shrinkage technique that incorporates the Hilbert transform, which induces shrinkage by pulling nearby eigenvalues together—making the shrinkage a local phenomenon and introducing essential nonlinearity into the estimation process.
In recent times, Ledoit and Wolf \cite{ledoit2021shrinkage,ledoit2022quadratic} developed an estimation technique that retains the eigenvectors of the sample covariance matrix but applies shrinkage to the inverse eigenvalues. This approach employs a quadratic shrinkage formula that involves two targets weighted by functions of the concentration ratio, enabling more refined covariance estimation in high-dimensional settings. An example of practical implementation in this domain is the \texttt{HDShOP} framework proposed by Bodnar et al. \cite{bodnar2025high}, which combines theoretical and empirical tools for high-dimensional shrinkage-based portfolio construction in \texttt{R}. Very recently, Tran et al. \cite{tran2025} introduced a multi-target shrinkage estimator, utilizing a grid search-based cross-validation method to determine optimal shrinkage intensities. Their approach demonstrated improved risk-adjusted returns and reduced volatility when tested on the Vietnamese stock market data during a stable financial period.

These developments underline the growing importance of combining robust statistical techniques with modern data and theory to overcome the limitations of classical MV portfolios. 

\subsection{Objective and Approach}

\textbf{Objective:} Given the theoretical underpinning and financial advantages of shrinkage estimators, this study seeks to examine and analyze MV and GMV models applying various shrinkage estimators of their model parameters, mean, and covariance matrix. We examine five shrinkage estimators for the mean and eleven for the covariance on six global markets, three of which are high-dimensional, $(p > n)$. The analysis employs a rolling window approach, using daily data that spans approximately eleven years for each market. Only eight covariance estimators have been found to be applicable in high-dimensional markets. Thus, there are sixty-six portfolios for the datasets where $n > p$, and forty-eight when $n < p$. This study also analyzes three out-of-sample window sizes: 3-months, 6-months, and 1-year, within the rolling window framework. Additionally, we categorize three groups: A, B, and C, based on the return-risk profile of investors, and perform a comprehensive analysis across groups. 

Overall, we conduct extensive empirical analysis to examine the advantages and performance of shrinkage estimators in portfolio selection.

\textbf{Sample Data:}  Daily closing prices of the constituents from the six global markets, namely, Dow Jones 30 (DJ 30), NIFTY 50, FTSE 100, S\&P 500, RUSSELL 1000, and TOPIX 1500, with the sample period of nearly 11 years from September 20, 2012, to Jun 6, 2024. 

\textbf{Execution:} Three of the six datasets, S\&P 500, RUSSELL 1000, and TOPIX 1500, are in the high-dimensional regime. In total, 48 MV and GMV-based portfolios are evaluated for high-dimensional datasets. For smaller-sized markets, DJ 30, NIFTY 50, and FTSE 100 (\(n > p\)), we solve 66 MV and GMV models. 

For the model computations, a rolling window scheme is executed for the in-sample (training) period of one year, and for the three different out-of-sample (testing) periods: (a) 3-months, (b) 6-months, and (c) 1-year. 

For comprehensive analysis, we consider three groups: A, B, and C, based on the risk-return profile of investors. The groups consist of investors focused on balancing return and risk metrics (Group A), aggressive investors prioritizing return and reward-risk ratios (Group B), and conservative risk-averse investors emphasizing risk mitigation (Group C).

We suggest utilizing a super-efficiency data envelopment analysis (DEA) model with portfolios as decision-making units (DMUs). The input and output for the DEA model are the various financial metrics, categorized by the three groups, and calculated over each out-of-sample period's return data.

\textbf{Identifying Best Shrinkage Portfolio:} For every market, for every group, and for each out-of-sample period, we arrange the portfolio efficiency scores in descending order and determine the top 10 portfolios. 
Subsequently, we select the two optimal portfolios by two approaches: (i) the market best performer and (ii) the universal best performer. A market best-performer identifies the top performers according to the underlying market, while a universally best performer endeavors to find the portfolio that excels across the widest range of markets. 

The market's top performance is the one that rates among the top 10 throughout all three out-of-sample periods and possesses the highest geometric mean of the three efficiency ratings.  The universal top performer is the portfolio that excels in the highest number of markets. 

\textbf{Comparative Analysis}—The set of market best performers and universally best performer portfolios for each category is compared with five traditional portfolio optimization models: the semi mean-absolute deviation (SMAD) model \cite{feinstein1993reformulation}, which builds upon the mean absolute deviation model \cite{konno1991}; the conditional value-at-risk model  \cite{rockafellar2000optimization}; the MiniMax model \cite{cai2000portfolio}; and the classical mean-variance (MV) and global minimum variance (GMV) models that utilize the sample mean and covariance matrix.  This comparative analysis relies on out-of-sample results for financial metrics and DEA super efficiency scores.

\textbf{Outcome:} The numerical study reveals multiple intriguing facts: (1) The choice of shrinkage estimators is more dependent on underlying market data, the choice is even more prudent for return-seeking investors, therefore, there is no single shrinkage estimator, which works well for all the markets and for all the types of investors. (2) When seeking the majority of occasions, the GMV model with Ledoit Wolf method of two parameter shrinkage covariance (COV2) is an optimal choice for wide range of investors while the MV model with COV2 together with sample mean (SM) is more appropriate for return-seeking investors. (3) Both the GMV+COV2 and COV2+SM models perform better than traditional benchmarks like classical Markowitz portfolios with sample estimates, MiniMax, CVaR, GMV, and SMAD. (4) The shrinkage covariance (COV2) frequently appears in top choices for several instances, the sample mean (SM) shines in the mean estimation category. (5) The sample mean together with the sample covariance results to poor out-of-sample results, rendering it an undesirable option.

Broadly, this study lays the groundwork for a more in-depth comprehension of how certain shrinkage models perform for a variety of investors and in various market scenarios.

\textbf{Structure:} The paper is organized as follows. Section \ref{sec2} presents the mean-variance and global minimum-variance portfolio optimization models. Section \ref{sec3} describes shrinkage estimators for the mean and covariance matrix. Section \ref{sec4} outlines the DEA super-efficiency model. Section \ref{sec5} explains the data and empirical setup. Section \ref{sec6} presents the results and discussion. Section \ref{sec7} offers a comparison with benchmark models. Finally, Section \ref{sec8} concludes by presenting our findings and outlining future research directions.

\section{Mean-Variance and Global Minimum-Variance Portfolio Optimization Models}\label{sec2}

Let \( \bm{x} \) represent a portfolio composed of \( p \) assets, denoted as \( \bm{x} = ( x_1, \dots, x_p )^{'} \), where \( x_i \) is a decision variable indicating the proportion of the total capital allocated to the \( i \)-th asset, for \( i = 1,  \dots, p \).  For $r_i$ be the random
return of the $i$th asset, 
the random vectors $R = (r_{1},\ldots,r_p)$ 
represent the random return vectors of $p$-assets
portfolio. We
then calculate the return of a portfolio $x$ as $R_x =
\sum_{i=1}^{p}r_{i}x_i$. For $E(r_i) = \mu_i\; i=1,\ldots,p$, the respective mean vector $\bm\mu$ and covariance matrix $\Sigma$ of a portfolio $x$ is expressed as $\bm\mu = (\mu_{1},\ldots,\mu_{p})$ and \( \Sigma = [\sigma_{ik}]_{p\times p}\) where $\sigma_{ik}= \textrm{covariance}(r_{i},r_{k}),\; i,\,k=1,\ldots,p$. For a given set of admissible
portfolios, $X = \{(x_1,\ldots,x_p)^{'},\;\; \sum_{i=1}^{p}x_i = 1, \; x_i \geq 0, \;i=1,\ldots,p\}$, where the nonnegativity constraints
$x_i \geq 0$ prohibit short-selling. The \textit{mean-variance} (MV) model is given by the following convex quadratic program.
\begin{align}
\text{(MV)} \quad  \min_{\bm{x} \in X} \quad & \bm{x}^\top \bm\mu - \gamma \bm{x}^\top\bm{\Sigma} \bm{x},
\nonumber
\end{align}
where \(\gamma > 0\) is a risk aversion parameter, varying it yields many efficient portfolios, each corresponding to a different risk-return profile of an investor. The \textit{global minimum-variance} model is described by
\begin{align}
\text{(GMV)} \quad  \min_{\bm{x} \in X} \quad & \bm{x}^\top\bm{\Sigma} \bm{x}.\nonumber
\end{align}

Traditionally, while solving the models (MV) or (GMV), the unknown parameters $\bm\mu$ and $\bm\Sigma$ are replaced by
their sample estimations using historical data or simulated data (or some empirical distribution such as uniform distribution). This results in unstable and non-robust portfolio weights. 
Shrinkage estimation emerges as a robust statistical tool that addresses these issues by systematically combining sample estimates with structured targets.
Shrinkage methods are based on the principle of trade-off between observed, data-driven estimates and a chosen target (often with desirable properties such as stability or sparsity) to minimize mean squared error.

\section{Shrinkage Estimators}\label{sec3}

\textit{Shrinkage method} is a linear combination of the sample
estimator of an underlying parameter and its another estimator. Formally, the shrinkage estimator for a parameter \(\bm{\theta}\), named $\hat{\bm{\theta}}_{shrink}$, can be expressed as:
\begin{equation}
\hat{\bm{\theta}}_{shrink} = \lambda\, \hat{\bm{\theta}}_{target} + (1 - \lambda)\, \hat{\bm{\theta}}_{sample},\nonumber
\end{equation}

where \( \hat{\bm{\theta}}_{sample} \) and \( \hat{\bm{\theta}}_{target} \) respectively, are the unbiased sample and target estimators of  \(\bm{\theta}\) 
and \( \lambda \in [0,1] \) is the shrinkage intensity parameter. 

\subsection{Shrinkage Estimator for Mean}
Shrinkage approaches to blend the sample mean vector with a more stable reference/target vector, such as a constant vector or the grand mean vector. This reduces sensitivity to outliers and estimation noise, leading to more robust mean inputs for the MV optimization model. The following describes four types of the most popular shrinkage estimators for the mean vector. 
\begin{itemize}
\item [1.] \textbf{The James-Stein shrinkage estimator (JS):}  Introduced by James and Stein \cite{james1961estimation}, it is a classical shrinkage approach that improves the estimation accuracy of the mean vector $\mu$ by shrinking the sample mean vector  \(\bar{\bm{r}}\) towards a common target. The JS estimator for the mean return vector is given by  
\begin{equation}
\hat{\bm{\mu}}_{JS} = \hat{\alpha} \hat{r}_0 \bm{1}_p+\left(1 - \hat{\alpha}\right)\bar{\bm{r}},   
 \label{JS1}
\end{equation}
where \(\bm{1}_p\) is a 
\(p\)-dimensional vector of ones, \( \hat{\alpha} \) is the shrinkage intensity and \(\hat{r}_0\bm{1}_p\) is the shrinkage target defined as:
\begin{equation}
\hat{r}_0 = \frac{\bar{\bm{r}}^\top \bm{S}_n^{-1} \bm{1}_p}{\bm{1}_p^\top \bm{S}_n^{-1} \bm{1}_p},\label{JStar}
\end{equation}
where \(\bm{S}_n\) denotes the sample covariance computed using sample size of $n$ and the shrinkage intensity is defined as
\begin{equation}
\hat{\alpha} = \min\left(1,\ \frac{p - 2}{n(\bar{\bm{r}} - \hat{r}_0 \bm{1}_p)^\top \bm{S}_n^{-1} (\bar{\bm{r}} - \hat{r}_0 \bm{1}_p)}\right).\nonumber
\end{equation}
\item [2.] \textbf{The Bayes-Stein shrinkage estimator (BS):}  Introduced in \cite{jorion1986bayes}, it arises from a Bayesian framework in which the true mean return vector is modelled as a random variable with a prior distribution. It represents a Bayesian improvement over the JS estimator by incorporating prior beliefs about the mean vector $\bm\mu$ and allowing a probabilistic interpretation of the shrinkage process. With the shrinkage target as given in equation (\ref{JStar}), the shrinkage intensity in equation (\ref{JS1}) for the BS estimator, is defined as:
\begin{equation}
\hat{\alpha} = \frac{p + 2}{p + 2 + n \left( \bar{\bm{r}} - \hat{r}_0\, \bm{1}_p \right)^\top \bm{S}_n^{-1} \left( \bar{\bm{r}} - \hat{r}_0\, \bm{1}_p \right)}\nonumber
\end{equation}


\item [3.] \textbf{The Quadratic shrinkage estimator (QUAD):} The QUAD estimator introduced by \cite{wang2014non}, proposes a data-driven shrinkage approach designed to estimate mean vector $\bm\mu$ under quadratic loss functions with unknown covariance matrices. The estimator is defined as:
\begin{equation}
    \hat{\bm{\mu}}_{Quad} = \frac{R_{2,n}R_{4,n}}{R_{1,n} + R_{2,n}-R_{3,n}} \mathbf{1}_p+\frac{R_{1,n} - R_{3,n}}{R_{1,n} + R_{2,n}-R_{3,n}} \, \bar{\bm{r}}
\end{equation}

where the coefficients \( R_{1,n}, R_{2,n}, R_{3,n}, R_{4,n} \) are defined as follows:
\begin{align*}
R_{1,n} &= \frac{1}{p(n-1)} \sum_{i \ne j} \bm{R}_{i}^\top \bm{S}_n^+ \bm{R}_{j}, &
R_{2,n} &= \frac{1}{np} \left( \sum_{k=1}^{n} \bm{R}_{k}^\top \bm{S}_n^+ \bm{R}_{k} - \frac{1}{n-1} \sum_{i \ne j} \bm{R}_{i}^\top \bm{S}_n^+ \bm{R}_{j} \right), \\ \nonumber
R_{3,n} &= \frac{1}{n\, \mathbf{1}_n^\top \bm{S}_n^+ \mathbf{1}_n} \sum_{k=1}^{n} \mathbf{1}_n^\top \bm{S}_n^+ \bm{R}_{k}, &
R_{4,n} &= \frac{1}{p(n-1)\, \mathbf{1}_n^\top \bm{S}_n^+ \mathbf{1}_n} \sum_{i \ne j} \mathbf{1}_n^\top \bm{S}_n^+ \bm{R}_{i} \bm{R}_{j}^\top \bm{S}_n^+ \mathbf{1}_n.\nonumber
\end{align*}
where \(\bm{S}_n^+\) is denoted for the Moore-Penrose inverse of \(\bm{S}_n\) and \(\bm{R}\) is the return rate data matrix of size \(n\times p\).

\item [4.] \textbf{The Bodnar optimal linear shrinkage estimator (BOP):} The BOP estimator is introduced by \cite{bodnar2019optimal} and is given by
\begin{equation}
\hat{\bm{\mu}}_{BOP} =  \hat{\alpha}\,\bm{\mu}_0+\hat{\beta}\,\bar{\bm{r}}.
\end{equation}
The shrinkage coefficients $\hat\alpha$ and $\hat\beta$ are obtained by minimizing the quadratic loss, which are almost surely asymptotically equivalent to 
\begin{equation}
\hat{\alpha} = (1 - \hat{\beta}) 
\frac{ \bar{\bm{r}}^{\top} \bm{S}_n^{-1} \bm{\mu}_0 }{ \bm{\mu}_0^{\top} \bm{S}_n^{-1} \bm{\mu}_0 }
, \quad
\hat{\beta} = \frac{
    \left(\bar{\bm{r}}^{\top} \bm{S}_n^{-1} \bar{\bm{r}} -\frac{c}{1 - c}\right)
     \bm{\mu}_0^{\top} \bm{S}_n^{-1} \bm{\mu}_0 - \left(\bar{\bm{r}}^{\top} \bm{S}_n^{-1} \bm{\mu}_0\right)^2 
}{
    \bar{\bm{r}}^{\top} \bm{S}_n^{-1} \bar{\bm{r}} \bm{\mu}_0^{\top} \bm{S}_n^{-1} \bm{\mu}_0 - \left( \bar{\bm{r}}^{\top} \bm{S}_n^{-1} \bm{\mu}_0 \right)^2
}\nonumber
\end{equation}
where \(c=p/n\) is the concentration ratio. In this study predetermined shrinkage target $\bm{\mu}_0$ is taken as $\bm{\mu}_0 = n^{\frac{\epsilon - 1}{2}}\mathbf{1}_p$, with $\epsilon \in (0,1)$. This estimator is shown to dominate other shrinkage methods in high-dimensional settings in terms of quadratic loss in both empirical and simulation studies \cite{bodnar2019optimal}.
\end{itemize}

\subsection{Shrinkage Estimators for Covariance Matrix}
Shrinkage covariance estimation becomes an essential tool in variance-based portfolio optimization models, especially when historical data is limited relative to the number of assets. Even when the number of data points is close to the number of assets, the classical sample covariance matrix becomes singular and hence is not a consistent estimator of the population covariance matrix. In high-dimensional settings, the shrinkage estimator of covariance plays a critical role in enhancing risk-adjusted performance and mitigating the pitfalls of traditional sample estimators \cite{bodnar2025high}. Below are the shrinkage estimators given for the covariance matrix: 


\begin{itemize}
\item[1.]  \textbf{Bodnar Linear Shrinkage (LS)}: 
This estimator is proposed by Bodnar et al. \cite{bodnar2014strong} and derived by minimizing the following expected quadratic loss  under the Frobenius norm:
\[
R(\bm{\Sigma}_{LS}) = \mathbb{E} \left( \left\| \bm{S}_n - \bm\Sigma_{LS} \right\|_F^2 \right),
\]
where 
$\bm{\Sigma}_{LS}$ is a LS estimator for covariance matrix, and  \(\| \cdot \|_F\) denotes the Frobenius norm. This approach yields an estimator that is a linear combination of the sample covariance matrix \(\bm{S}_n\) and a predetermined target matrix \(\bm{\Sigma_0}\). Further discussion on the selection of \(\bm{\Sigma_0}\) can be found in \cite{ledoit2004well, bai2011estimating}.
The LS estimator is derived from above as: 
\begin{equation}
\widehat{\bm{\Sigma}}_{LS} = \hat{\alpha}\, \bm{S}_n + \hat{\beta}\, \bm{\Sigma}_0 \\
\end{equation}
where the shrinkage coefficients \(\hat{\alpha}\) and \(\hat{\beta}\) are computed as: 
\[
\hat{\alpha} = 1 - \frac{1}{n} \frac{\| \bm{S}_n \|_{tr}^2 \, \| \bm{\Sigma}_0 \|_F^2}{\| \bm{S}_n \|_F^2 \, \| \bm{\Sigma}_0 \|_F^2 - \| \bm{S}_n \bm{\Sigma}_0 \|_{tr}^2},\quad
\hat{\beta} = \frac{\| \bm{S}_n \bm{\Sigma}_0 \|_{tr}}{\| \bm{\Sigma}_0 \|_F^2} \left( 1 - \hat{\alpha} \right).
\]

Here, \(\| \cdot \|_{\mathrm{tr}}\) denotes the trace norm. By construction, the estimator satisfies $\hat\alpha \leq 1$ and $\hat\beta \geq 0$.
The LS estimator incorporates prior structure through the target matrix \( \bm{\Sigma}_0 \), which can enhance performance when prior information is informative. In this study, we take $\bm\Sigma_0 = \frac{1}{p}\bm{I}_p$ \cite{bodnar2014strong}. 

\item [2.] \textbf{Ledoit Wolf Method (LW):} 
Ledoit and Wolf \cite{ledoit2003improved,ledoit2004well} introduce a family of covariance shrinkage estimators that improve the performance of the sample covariance matrix by shrinking it toward well-conditioned, structured targets. These target matrices include the identity matrix, the combination of the identity matrix and a matrix of ones, the constant correlation matrix, the diagonal matrix of variances, and a single-factor market model, which are referred to as COV1, COV2, COVCOR, COVDIAG, and COVMKT, respectively.  
The LW shrinkage estimator is defined as: 
\begin{equation}
\widehat{\bm\Sigma}_{LW} = \hat{\lambda} \bm{T} + (1 - \hat{\lambda}) \bm{S}_n,\label{LW}
\end{equation}
where, $\bm{T}$ is the target matrix. By minimizing the mean-squared error between the estimator and the true covariance matrix, the optimal shrinkage intensity is obtained as
\begin{equation}
    \hat{\lambda} = \min\left(1, \max\left(0, \frac{\hat{\pi} - \hat{\rho}}{n\hat{\gamma}} \right)\right).\label{lambda}
\end{equation}

Here, $\hat{\pi}$ and \(\hat{\gamma}\) are given by: 
\[
\hat{\pi} = \sum_{i=1}^p \sum_{j=1}^p \left( (S_n^{(2)})_{ij} - (S_n)_{ij}^2 \right), \;\; \hat{\gamma} = \sum_{i=1}^p \sum_{j=1}^p \left( (S_n)_{ij} - T_{ij} \right)^2
\]

where the second-moment matrix is  \( \bm{S}_n^{(2)} = \frac{1}{n} (\bm{R}^{(2)})^\top \bm{R}^{(2)}, \) with \((\bm{R}^{(2)})_{ij} = \bm{R}_{ij}^2\) and 
\begin{equation}
    \hat{\rho} = \hat{\rho}_{\mathrm{diag}} + \hat{\rho}_{\mathrm{off}},\label{pho}
\end{equation}
with the diagonal contribution is specified as: $\hat{\rho}_{\mathrm{diag}} = \sum_{i=1}^p \left( (S_n^{(2)})_{ii} - (S_n)_{ii}^2 \right)$ while the off-diagonal component \(\hat{\rho}_{\mathrm{off}}\) varies from estimator to estimator.

\begin{enumerate}
    \item[(2a)] \textbf{One-parameter shrinkage covariance estimator (COV1)}~\cite{ledoit2003improved}: This estimator sets \(\bm{T} = \bar{v} \bm{I}_p\), where \(\bar{v} = \frac{1}{p} \mathrm{trace}(\bm{S}_n)\), and uses \(\hat{\rho} = 0\) in \eqref{lambda}.


\item[(2b)] \textbf{Two-parameter shrinkage covariance estimator (COV2)}~\cite{ledoit2003improved}: This estimator sets \(\bm{T} = \bar{v} \bm{I}_p + \bar{c} (\bm{J}_p - \bm{I}_p)\), where \(\bar{v} = \frac{1}{p} \mathrm{trace}(\bm{S}_n)\), and \(\bm{J}_p\) is the \(p \times p\) matrix of ones. It uses
\[
\hat{\rho}_{\mathrm{off}} = \frac{1}{p - 1} \left[ \frac{1}{pn} \sum_{t=1}^n \left( \left( \sum_{i=1}^p R_{ti} \right)^2 - \sum_{i=1}^p R_{ti}^2 \right)^2 - \frac{1}{p} \left( \sum_{i \neq j} (S_n)_{ij} \right)^2 \right]
\]
in \eqref{pho}.



\item[(2c)] \textbf{Constant-correlation shrinkage covariance estimator (COVCOR)~\cite{ledoit2003improved}:} This estimator sets 
\(\bm{T}\) defined elementwise by
 \[
 T_{ij} = 
 \begin{cases}
 \bar{r} \sqrt{(S_n)_{ii}} \sqrt{(S_n)_{jj}}, & \text{if } i \ne j, \\
 (S_n)_{ii}, & \text{if } i = j,
 \end{cases}
 \]
 where $\bar{r}$ is computed as $\bar{r} = \frac{1}{p(p-1)} \sum_{i \neq j} \frac{(S_n)_{ij}}{\sqrt{(S_n)_{ii}(S_n)_{jj}}}.$ This estimator uses the off-diagonal shrinkage adjustment in \eqref{pho} as
\[
 \hat{\rho}_{\mathrm{off}} = \bar{r} \sum_{i=1}^p \sum_{j=1}^p \frac{\Gamma_{ij}}{(S_n)_{ii}(S_n)_{jj}},
\]
where $\bm{\Gamma} = \frac{1}{n} (\bm{R}^{(3)})^\top \bm{R} - \left( \bm{v} \cdot \bm{1}_p^\top \right) \circ \bm{S}_n,$
with \((\bm{R}^{(3)})_{ij} = R_{ij}^3\), \(\bm{v} = \mathrm{diag}(\bm{S}_n)\), and \(\circ\) denotes Hadamard product.


\item [(2d)] \textbf{Diagonal shrinkage covariance estimator (COVDIAG)~\cite{ledoit2004well}:}  This estimator sets \(\bm{T} = \mathrm{diag}(\bm{S}_n)\),  and uses \(\hat{\rho} = \hat{\rho}_{diag}\) in \eqref{lambda}.

\item[(2e)] \textbf{Market shrinkage covariance estimator (COVMKT)~\cite{ledoit2004well}:} This estimator sets 
\(\bm{T}\) defined elementwise by
\[
T_{ij} = 
\begin{cases}
\displaystyle\frac{(\bm{\Sigma}_{R R_{\mathrm{mkt}}})_i (\bm{\Sigma}_{R R_{\mathrm{mkt}}})_j}{\sigma_{R_{\mathrm{mkt}} R_{\mathrm{mkt}}}}, & \text{if } i \ne j, \\[0.5em]
(S_n)_{ii}, & \text{if } i = j.
\end{cases}
\]
where \(\bm{\Sigma}_{R R_{\mathrm{mkt}}} \text{ and } \sigma_{R_{\mathrm{mkt}} R_{\mathrm{mkt}}}\) are defined as
\[
\bm{\Sigma}_{R R_{\mathrm{mkt}}} = \frac{1}{n} \bm{R}^\top \bm{R}_{\mathrm{mkt}}, \; \sigma_{R_{\mathrm{mkt}} R_{\mathrm{mkt}}} = \frac{1}{n} \bm{R}_{\mathrm{mkt}}^\top \bm{R}_{\mathrm{mkt}} \; \textrm{where} \; \bm{R}_{\mathrm{mkt}} = \frac{1}{p} \bm{R} \bm{1}_p.
\]

Also, this estimator uses the off-diagonal shrinkage adjustment in \eqref{pho} as
\[
\hat{\rho}_{\mathrm{off}} = 2 \sum_{i=1}^p \sum_{j=1}^p \frac{(\bm{K}_1)_{ij} (\bm{\Sigma}_{R R_{\mathrm{mkt}}})_j}{\sigma_{R_{\mathrm{mkt}} R_{\mathrm{mkt}}}} 
- \sum_{i=1}^p \sum_{j=1}^p \frac{(\bm{K}_2)_{ij} (\bm{\Sigma}_{R R_{\mathrm{mkt}}})_i (\bm{\Sigma}_{R R_{\mathrm{mkt}}})_j}{\sigma_{R_{\mathrm{mkt}} R_{\mathrm{mkt}}}^2} 
- \sum_{i=1}^p \frac{(\bm{K}_1)_{ii} (\bm{\Sigma}_{R R_{\mathrm{mkt}}})_i^2}{\sigma_{R_{\mathrm{mkt}} R_{\mathrm{mkt}}}^2},
\]
where
\[
\bm{K}_1 = \frac{1}{n} (\bm{R}^{(2)})^\top \bm{M} - \bm{\Sigma}_{R R_{\mathrm{mkt}}} \circ \bm{S}_n, \; \bm{K}_2 = \frac{1}{n} \bm{M}^\top \bm{M} - \sigma_{R_{\mathrm{mkt}} R_{\mathrm{mkt}}} \bm{S}_n,\;\; \textrm{with}\; \bm{M} = \bm{R} \circ \bm{R}_{\mathrm{mkt}}^\top
\]




\end{enumerate}
    \item[3.] \textbf{Inverse Shrinkage-Based variance Estimators:} Building on the foundational work of Stein~\cite{stein1975estimation, stein1977estimation, stein1986lectures}, recent developments reinterpret and refine his original shrinkage approach to improve both theoretical tractability and practical applicability in high-dimensional settings \cite{ledoit2021shrinkage,ledoit2022quadratic}. Specifically, this class of estimators transforms the nonlinear shrinkage of eigenvalues into a linear shrinkage in the inverse-eigenvalue domain.
Let the sample covariance matrix \(\bm{S}_n\) admits the spectral decomposition
\[
\bm{S}_n = \bm{U} \bm{\Lambda} \bm{U}^\top,
\]
where \(\bm{\Lambda} = \mathrm{diag}(\lambda_1, \dots, \lambda_p)\) contains the eigenvalues arranged in increasing order, and \(\bm{U} = [u_1, \dots, u_p]\) is the corresponding matrix of orthonormal eigenvectors. The key idea is to 
shrink the inverse eigenvalues toward structured targets, and then invert back to obtain stabilized covariance estimates. Below are the three estimators under this framework. 

\begin{enumerate}
    \item[(3a)] \textbf{Linear Inverse Shrinkage Estimator (LIS):} Proposed by Ledoit and Wolf~\cite{ledoit2021shrinkage}, it modifies the inverse eigenvalues of the sample covariance matrix using a linear shrinkage approach. The LIS inverse eigenvalues are defined as 
\[
\widehat{\delta}_i^{\mathrm{LIS}} = (1 - c) x_i + 2c x_i \tilde{\theta}_n(x_i),\text{ for }1\leq i\leq p.
\]
where $c\in(0,\infty)$ is concentration ratio,
\(
x_i = \lambda_i^{-1},
\)
which correspond to the nonzero sample eigenvalues and the smoothed Stein shrinkage function \(\tilde{\theta}_n(x_i)\) is given by
$
\tilde{\theta}_n(x_i) = \frac{1}{p} \sum_{j=1}^{\min(p,n)} \frac{x_j (x_j - x_i)}{(x_j - x_i)^2 + h_n^2 x_j^2}$  with the smoothing parameter \( h_n \) in this study is taken as \(
h_n = \frac{\min\;(c^2, \;,1/c^2)^{0.35}}{p^{0.35}}.
\)

To ensure regularization from above, the inverse eigenvalues are truncated as: $\delta_i^{\mathrm{LIS}} = \min(x_i, \widehat{\delta}_i^{\mathrm{LIS}})$. Therefore, with \( \bm\Delta^{\mathrm{LIS}} = \mathrm{diag}(1/\delta_1^{\mathrm{LIS}}, \dots, 1/\delta_p^{\mathrm{LIS}}) \), the LIS estimator of the covariance matrix is given as
\[
\widehat{\bm\Sigma}_{LIS} = \bm{U} \Delta^{LIS} \bm{U}^\top.
\]

\item [(3b)] \textbf{Quadratic Inverse Shrinkage (QIS):} In this case, Ledoit and Wolf~\cite{ledoit2022quadratic}, modifies the inverse eigenvalues of the sample covariance matrix using a quadratic shrinkage approach. For \( p \leq n \), the QIS shrunk eigenvalues are computed as
\[
\delta_i = \frac{1}{(1 - c)^2 x_i + 2c(1 - c)x_i \tilde{\theta}_n(x_i) + c^2 x_i A^2_{\theta}(x_i)},\text{ for }1\leq i\leq p.
\]
where $ A^2_{\theta}(x_i) = \tilde{\theta}_n(x_i)^2 + \left(\frac{1}{p} \sum_{j=1}^{\min(p, n)} \frac{h_n x_j^2}{(x_j - x_i)^2 + h_n^2 x_j^2}\right)^2
$. While for \( p > n \), the shrinkage for null eigenvalues is given as
\[
\delta_0 = \frac{1}{(c - 1) \cdot \frac{1}{p - n} \sum_{j=1}^{p - n} x_j},
\]
and set
\[
\delta_i =
\begin{cases}
\delta_0, & \text{for } i = 1, \ldots, p - n, \\
\displaystyle\frac{1}{x_i A^2_{\theta}(x_i)}, & \text{for } i = p - n + 1, \ldots, p.
\end{cases}
\]

To ensure trace preservation, define the normalized eigenvalues as $
\delta_i^{\mathrm{QIS}} = \delta_i \left( \frac{\sum_{i=1}^{p} \lambda_{n,i}}{\sum_{i=1}^{p} \delta_i} \right).$ With \( \bm\Delta^{\mathrm{QIS}} = \mathrm{diag}(\delta_1^{\mathrm{QIS}}, \ldots, \delta_p^{\mathrm{QIS}}) \), the QIS covariance estimator is given as
\[
\widehat{\bm{\Sigma}}_{QIS} = \bm{U}  \bm{\Delta}^{\mathrm{QIS}}  \bm{U}^\top.
\]

\item[(3c)] \textbf{Geometric Inverse Shrinkage Estimator (GIS):} This method combines LIS and QIS by applying geometric averaging on the inverse eigenvalues (Ledoit and Wolf~\cite{ledoit2022quadratic}) 
\[
\delta_i^{\mathrm{GIS}} = \sqrt{\frac{\delta_i^{\mathrm{QIS}}}{\delta_i^{\mathrm{LIS}}}}, \text{ for } 1\leq i\leq p.
\]
For \( \bm\Delta^{\mathrm{GIS}} = \mathrm{diag}(\delta_1^{\mathrm{GIS}}, \ldots, \delta_p^{\mathrm{GIS}}) \), the GIS estimator of the covariance matrix is defined as
\[
\widehat{\bm{\Sigma}}_{GIS} = \bm{U} \bm\Delta^{\mathrm{GIS}} \bm{U}^\top.
\]
\end{enumerate}

\item [4.] \textbf{Analytic Shrinkage Estimator (AS)}: Introduced by Ledoit and Wolf~\cite{ledoit2020analytical}, it uses kernel-based estimation of the spectral density to derive nonlinear shrinkage eigenvalues analytically.
To estimate the spectral density, select a global bandwidth \( h_n = n^{-1/3} \), and define the locally adaptive bandwidth as \( h_{n,j} = h_n \lambda_j, \forall j=1,\ldots ,n\). Let \( \kappa_{ij} = \frac{\lambda_i - \lambda_j}{h_{n,j}} \).
The Epanechnikov kernel estimate of the spectral density at eigenvalue $\lambda_{n,i}$ is given by
\[
\widehat{f}_n(\lambda_{i}) = \frac{1}{p} \sum_{j=1}^{p} \frac{3}{4 \sqrt{5} h_{n,j}} \left( 1 - \frac{\kappa_{ij}^2}{5}  \right)_+, \text{where } [x]_+ = \max\{x, 0\}.
\]
The Hilbert transform of the spectral density is computed as
\[
\tilde{H}_{\widehat{f}_n}(\lambda_i) = \frac{1}{p} \sum_{j=1}^{p} \frac{1}{h_{n,j} \pi} \left[ -\frac{3}{10}\kappa_{ij} + \frac{3}{4\sqrt{5}} \left(1 - \frac{\kappa_{ij}^2}{5}\right) \log \left| \frac{\sqrt{5} - \kappa_{ij}}{\sqrt{5} + \kappa_{ij}} \right| \right].
\]
and the Analytic Shrinkage (AS) eigenvalues are then computed as
\[
\delta_i^{\mathrm{AS}} = \frac{\lambda_i}{\left[ \pi c \lambda_i \tilde{f}_n(\lambda_i) \right]^2 + \left[ 1 - c - \pi c \lambda_i \tilde{H}_{\widehat{f}_n}(\lambda_i) \right]^2}.
\]
For \( \bm\Delta^{\mathrm{AS}} = \mathrm{diag}(\delta_1^{\mathrm{AS}}, \ldots, \delta_P^{\mathrm{AS}}) \), the AS covariance estimator is defined as
\[
\widehat{\bm{\Sigma}}_{AS} = \bm{U} \bm\Delta^{\mathrm{AS}} \bm{U}^\top.
\]
\end{itemize}

Table \ref{listing} summarizes the shrinkage estimators considered in this paper for easy reference. 
\begin{table}[ht]
\centering
\footnotesize
\caption{A total of five shrinkage estimators are implemented for the mean vector and eleven for the covariance matrix. Out of the eleven covariance estimators, GIS, LIS, and AS are not applicable for the high-dimensional data sets, i.~e, when the number of time points exceeds the number of assets ($p > n$).}
\begin{tabular}{|l|l|}
\hline
\textbf{Mean vector} & \textbf{Covariance matrix} \\\hline
1. Classical approach: Sample mean (SM)  &   1. Classical approach Sample covariance (SCV)\\ 
2. James-Stein shrinkage (JS) \cite{james1961estimation} & 2. Linear shrinkage (LS) \cite{bodnar2014strong}\\
3. Bayes-Stein shrinkage (BS) \cite{jorion1986bayes}  & 3. One parameter shrinkage covariance (COV1) \cite{ledoit2003improved}  \\ 
4. Bodnar optimal linear shrinkage (BOP) \cite{bodnar2019optimal}  & 4. Two parameter shrinkage covariance  (COV2) \cite{ledoit2003improved}  \\
5. Quadratic shrinkage (QUAD) \cite{wang2014non}  & 5. Constant-correlation shrinkage covariance (COVCOR) \cite{ledoit2003improved}      \\
  & 6. Diagonal shrinkage covariance (COVDIAG) \cite{ledoit2004well} \\ 
 & 7. Market shrinkage covariance (COVMKT) \cite{ledoit2004well} \\ 
 & 8. Linear Inverse Shrinkage (LIS) \cite{ledoit2021shrinkage} \\ 
 & 9. Quadratic Inverse Shrinkage (QIS) \cite{ledoit2022quadratic}  \\ 
 & 10. Geometric Inverse Shrinkage (GIS) \cite{ledoit2022quadratic}  \\ 
 &  11. Analytic shrinkage (AS) \cite{ledoit2020analytical} \\ 
\hline
\end{tabular}

\label{listing}
\end{table}

For the five estimators of the mean vector combined with the eleven covariance matrix estimators, we construct a total of 55 (5 × 11) MV models and 11 GMV models in the low-dimensional setting. In the high-dimensional case, using only eight covariance matrix estimators, we obtain 40 (5 × 8) MV models and 8 GMV models. With three out-of-sample-sized windows, resulting in a 
large number of portfolios. 

We set $\gamma = 1$ in the (MV) model in our empirical analysis.

We need an effective and robust ranking method that incorporates key financial performance metrics, such as returns, risk indicators, and risk–reward ratios. We deploy the Data Envelopment Analysis (DEA) super-efficiency model~\cite{charnes1978measuring} to rank the portfolios generated from the shrinkage-based MV and GMV models.


\section{Super-efficiency DEA Model}\label{sec4}
DEA is a non-parametric linear programming approach used to evaluate the relative efficiency of homogeneous decision-making units (DMUs) that operate with multiple inputs and outputs. In the standard DEA framework, an efficient frontier is constructed based on the best-performing units. However, since all efficient units receive the same efficiency score, Andersen and Petersen~\cite{andersen1993procedure} introduced the super-efficiency DEA model to enable a complete ranking.
In the super-efficiency model, the DMU under evaluation is excluded from the reference set, allowing for efficiency scores greater than one for efficient DMUs and enabling a more granular ranking among them. The super-efficiency DEA model is well-suited for benchmarking where discrimination among efficient units is essential. 

Let there be \( d \) units, each utilizing \( m \) inputs to produce \( s \) outputs. Define the input matrix \( \bm{W} \in \mathbb{R}^{d \times m} \), where each row corresponds to a DMU and each column to an input variable. Similarly, let the output matrix \( \bm{Y} \in \mathbb{R}^{d \times s} \) represent the outputs of all DMUs. The input and output vectors for DMU \( i \) are denoted \( \bm{w}_i \in \mathbb{R}^m \) and \( \bm{y}_i \in \mathbb{R}^s \), respectively. The super-efficiency score of DMU \( i \), denoted \( \textit{eff}_i \),\, $i=1,\dots,d,$ is obtained by solving the following optimization problem:

\begin{align*} \text{(DEA)}\;\; \max \quad & \textit{eff}_i= \sum_{r=1}^s u_r y_{ir} \\ \text{subject to} \quad & \sum_{r=1}^s u_r y_{jr} - \sum_{k=1}^m v_k w_{jk} \leq 0, \quad \forall\; j \neq i,\; j=1,\ldots, d, \\ & \sum_{k=1}^m v_k w_{ik} = 1,\\ & u_r \geq 0, \quad v_k \geq 0, \quad \forall\; r, \,k. \end{align*}
Here, \( u_r \) and \( v_k \) denote the weights assigned to output \( r \) and input \( k \), respectively.

In the context of this paper, we consider portfolios generated by various shrinkage-based MV and GMV models as DMUs. The objective is to rank these units based on multiple financial performance metrics summarized in Table \ref{finlisting}. Input metrics (to be minimized) are primarily associated with risk measures, while output metrics (to be maximized) represent return and risk-adjusted ratios. The (DEA) model provides a systematic framework to evaluate shrinkage strategies by assessing how effectively they manage the trade-off between return and risk.

\begin{table*}[ht]
\centering
\caption{Financial metrics used for computing the super-efficiency scores and comparative analysis with the benchmark models. }
\scalebox{.85}{%
\begin{tabular}{p{4.5cm}| p{2cm} | p{6cm} }
\hline
\textbf{Metric} & \textbf{Input/Output} & \textbf{Formula / Description} \\
\hline
\textbf{Return} & & \\
\quad Mean return  & Output & $E(R_x) = \displaystyle \frac{1}{n} \sum_{l=1}^{n} R_l$; $R_l$ is a out-of-sample return of portfolio $x$ at time $l; l=1,\ldots,n$ \\
\hline
\textbf{Risk} & & \\
\quad Standard Deviation (SD) & Input & $ \sqrt{\frac{1}{n} \sum_{l=1}^{n} \left(E(R_x) - R_l\right)^2}$\, 
\\
\quad Value-at-Risk (VaR$_{0.05}(R_x)$) & Input & Out-of-sample Value-at-risk of portfolio $x$ at 95\% confidence level\footnote{Typically calculated using historical simulation or parametric methods.} \\
\quad Conditional VaR (CVaR$_{0.05}$) & Input & Out-of-sample conditional VaR of portfolio $x$ at 95\% confidence level\footnote{Also known as Expected Shortfall.} \\
\quad Downside Deviation (DD) & Input & $\sqrt{ \frac{1}{n} \sum_{l=1}^{n} \left( \min(0, R_l) \right)^2 }$ \\
\hline
\textbf{Risk-adjusted Performance} & & \\
\quad Mean-CVaR ratio & Output & $\displaystyle \frac{E(R_x)}{\text{CVaR}_{0.05}(R_x)}$;\, $E(R_x) > 0$ \\
& & \\
\quad Sharpe ratio & Output & $\displaystyle \frac{E(R_x)}{\text{SD}}$;\, $E(R_x) > 0$ \\
& & \\
\quad Sortino ratio & Output & $\displaystyle \frac{E(R_x)}{\text{DD}}$;\, $E(R_x) > 0$ \\
& & \\
\quad Mean-VaR ratio & Output & $\displaystyle \frac{E(R_x)}{\text{VaR}_{0.05}(R_x)}$;\, $E(R_x) > 0$ \\
& & \\
\quad Turnover & Input & $\displaystyle \frac{1}{n-1} \sum_{l=1}^{n-1} \sum_{j=1}^{p} \left|x_{j,l+1} - x_{j,l}\right|$ \\
\hline
\end{tabular}
}
\label{finlisting}
\end{table*}

\section{Data and Empirical Set-up}\label{sec5}
\subsection{Sample Data and Sample Period}
The sample data consists of daily closing prices of the constituents for the six global market indices listed in Table \ref{markets} from September 20, 2012, to Jun 6, 2024. Daily historical price data is collected from the LSEG workspace (https://www.lseg.com/en), previously Refinitiv Eikon datastream. Only assets that are actively trading and part of the index as of Jun 6, 2024.
\begin{table}[ht]
\centering
\footnotesize
\caption{Financial data sets for the empirical study}\label{markets}
\begin{tabular}{|l|l|}
\hline
\textbf{Low-dimensional data sets} & \textbf{High-dimensional Data Sets} \\\hline
\textbf{Data Set 1:} DJ 30 (USA), 30 assets & \textbf{Data Set 4:} S\&P 500 (USA), 453 assets \\
\textbf{Data Set 2:} NIFTY 50 (India), 48 assets & \textbf{Data Set 5:} RUSSELL 1000 (USA), 664 assets  \\
\textbf{Data Set 3:} FTSE 100 (UK), 92 assets & \textbf{Data Set 6:} TOPIX 1500 (Japan), 1451 assets \\
\hline
\end{tabular}

\end{table}

The experiments are performed on a Windows 10 64-bit operating system with 12 GB RAM and a 12th Gen Intel(R) Core(TM) i7-12700T 1.40 GHz processor. We use MATLAB YALMIP with the MOSEK solver to solve all considered optimization models.

\subsection{Rolling Window Scheme}
We use a rolling window approach consisting of an in-sample period and an  out-of-sample period,  measured in days. All optimal portfolios are obtained by solving corresponding models over the in-sample data, which are then assessed on the out-of-sample data.  The in-sample time horizon is then shifted forward by a number of days corresponding to the out-of-sample period to create a second window, and the previous procedure is repeated until the entire data set is exhausted. For the present study, we use a 1-year in-sample size (260 trading days) and three out-of-sample periods: \textbf{(1)} 3-months (65 trading days), leading to 43 windows (in-sample of one year is shifted forward by 3-months); \textbf{(2)} 6-months (130 trading days), leading to 21 windows; and \textbf{(3)} 1-year (260 trading days), resulting in 10 windows. 

For 260 days in-sample, the concentration ratio $c > 1$ for three high-dimensional datasets, the GIS, LIS, and AS shrinkage estimators for covariance, are not defined as they require $c < 1$. Therefore, we have 55 (5$\times$11) MV and 11 GMV models for low-dimensional data and 40 (5$\times$8) MV and 8 GMV models for high-dimensional data in each in-sample rolling window. Each optimal portfolio is assessed across three out-of-sample windows.

This analysis of varying out-of-sample window sizes aims to provide information on the stable performance of all considered models, particularly in relation to risk-return dynamics, risk-adjusted efficiency, and the stability of portfolio allocation with respect to changes in window size.

\subsection{Methodology}
The empirical investigation of the underlying study is mainly divided into the following steps:
\begin{itemize}
    \item Step I- \textbf{Ranking } We first rank all the portfolios generated from the shrinkage-based MV and GMV models by (DEA) models utilizing the input/output as described in Table \ref{finlisting}. For this, the ranking of portfolios is done for each market, each group, and for the three sized out-of-sample windows as follows:
\begin{enumerate}
    \item \textbf{Group A}: All ten metrics are taken into consideration. That means the input-output data matrix is $66\times 10$ (for a low-dimension market) and $48 \times 10$ (for a high-dimensional market), and we have three such matrices for each of the out-of-sample periods.
    
    \item \textbf{Group B}: When only return and risk-adjusted metrics are taken in analysis.     
    The model includes five outputs namely average return, mean-CVaR ratio, Sharpe ratio, Sortino ratio, and mean-VaR ratio, and the only input is set as a vector of ones, implying equal weight or neutral performance across all DMUs. Thus, the input-output data matrix is $66\times 6$ (for a low-dimensional market) and $48 \times 6$ (for a high-dimensional market).

    \item \textbf{Group C}: When only the risk-based financial metrics are applied in ranking. The inputs for this group are standard deviation (SD), value-at-risk (VaR), conditional value-at-risk (CVaR), maximum drawdown (DD), and turnover, capturing pure risk characteristics. The only output is taken as a vector of ones. Thus, the input-output data matrix is $66\times 6$ (for a low-dimensional market) and $48 \times 6$ (for a high-dimensional market).

\end{enumerate}

\item Step II- \textbf{Filtration:}  Based on their efficiency scores, we identify the top 10 portfolios for every market, every group, and each of the three out-of-sample windows. Out of these 10 most efficient portfolios, we finally select the two best portfolios for each group in two ways: (i) \textbf{market best-performer} and (ii) \textbf{universal best-performer}.

\begin{itemize}
 \item For each of the three groups, the market's best-performing portfolio is the one having the highest geometric mean (GM) $(eff_1\;eff_2\;eff_3)^{1/3}$ of the efficiency scores in the three out-of-sample windows. Thus, we have 6 of the best-performing market portfolios (one for each market) for every group.
 
 \item For each group, the universal best-performing portfolio is the one that is the best market performer in the maximum number of markets.

 \end{itemize}

\item Step III-\textbf{Comparison with the benchmark models:} 
The final set of two filtered portfolios for each group are  compared against the following traditional PO models over each data set:
\begin{itemize}
    \item [1.] Classical Markowitz mean-variance (MV) model with the sample mean and the sample covariance \cite{markowitz1952portfolio}.
    \item [2.] Classical global minimum variance (GMV) model with sample covariance \cite{markowitz1952portfolio}.  
    \item [3.] Semi Mean Absolute Deviation (SMAD) model \cite{feinstein1993reformulation}.
    \item [4.] Conditional Value at Risk (CVaR) model \cite{rockafellar2000optimization}.
    \item [5.] MiniMax (MM) model \cite{cai2000portfolio}. 
    \end{itemize}
The mathematical formulations of the optimization problems for the SMAD, CVaR, and MM models are provided in the Appendix.  
The comparative analysis provided a comprehensive evaluation of the optimal portfolios from the shrinkage-based PO with widely accepted benchmark models across the considered markets.

\end{itemize}

\section{Result Discussion}\label{sec6}
\subsection{Assessment for Group A }
The top 10 portfolio models are shown in Table \ref{tab1} based on their efficiency score computed by the (DEA) model when all financial metrics (listed in Table \ref{finlisting}) are taken into account as input and output. 

The models were listed in the format of \textbf{``covariance estimator + mean estimator"} for the (MV) model, while it is expressed as \textbf{``GMV + covariance estimator"} for the (GMV) model, indicating that the mean estimator has no role to play herein. 

Ranking is presented for the three distinct window sizes across each of the six datasets (Panels 1–3).  Panel 4 presents the market and universal best performers for each dataset. The efficiency score for each portfolio across all datasets and varying out-of-sample window sizes is presented in the GitHub link \footnote{https://github.com/rupendrayadav19401/Shrinkage-PO}.

The following conclusions are drawn from Table \ref{tab1}.
    \begin{itemize}
        \item \textit{DJ:}  GMV+COV2 was identified as the best-performing model, as it appears in the top 5 list over all window sizes, with the highest geometric mean efficiency score of 1.01043. We observe that COV2 demonstrates superior performance relative to other covariance estimators, followed by COV1. This is evidenced by the fact that, out of the thirty top 10 portfolios, 18 are based on COV2 and 9 on COV1. Notably, portfolios from the model GMV+COV2 appeared consistently across all three out-of-sample windows. This suggests that the COV2 becomes the best shrinkage estimator. 
        
          \item \textit{NIFTY:} GMV+COV2 demonstrates superior performance as a shrinkage model, achieving a geometric mean efficiency of 1.00512. GMV+LIS ranks as the second-best performer, maintaining a position within the top two selections for the three-month and one-year out-of-sample periods. Portfolios employing sample and BS estimators for mean estimation exhibited superior performance, specifically over the 6-month and 1-year out-of-sample periods, respectively.
          
\item \textit{FTSE:} LIS+SM is the top-performing shrinkage model, achieving a geometric mean efficiency of 1.00285. The GIS, LIS, and QIS covariance estimators outperformed the others, and the BS, sample, and QUAD estimators estimate the mean more accurately than the others. It's notable that GMV-based models don't make it to the top 10  rankings. 

\item \textit{S}\&\textit{P:} Like DJ and NIFTY, the GMV+COV2 model distinguishes itself as the foremost performer, achieving a geometric mean efficiency of 1.07442. Regardless of the out-of-sample duration, it consistently demonstrates exceptional performance in this market. Moreover, the COV2 and QIS shrinkage estimators of covariance exhibited superior performance during the 6-month out-of-sample interval.  During the 3-month and 1-year intervals, their efficacy was, however, less pronounced.
   
\item\textit{RUSSELL:} GMV+QIS and GMV+COV2 shrinkage-based portfolios exhibited comparable performance within this data set, with geometric means efficiency of 1.03598 and 1.0171, respectively. For all three out-of-sample periods, QIS+BS also retains its top 10 position for this market. In fact, portfolios utilizing the QIS variance estimator consistently ranked among the top 10 performers here. 

 \item \textit{TOPIX:}  GMV+COVMKT model is identified as the best-performing approach with the geometric mean efficiency value 1.05409, followed by the GMV+SCV model with a score of 1.02523. The GMV-based models consistently dominated the top 10 positions. The GMV+COV2 estimator retains its top 10 position over the 3-month and 6-month out-of-sample period; however, it missed for the 1-year period, with an efficiency score of 0.97899. Among non-GMV models, the COVMKT+BS model demonstrated superior performance.

\item \textit{Universal best performer:} Except for the FTSE and TOPIX (over the 1-year window), the GMV+COV2 combination consistently placed in the top 10 across all datasets and window types.  This renders the portfolio a universally superior performer. The S\&P, a globally recognized market, consistently ranks GMV-COV2 as the top model across all window widths.

\item \textit{Observations on varying out-of-sample period:} When the out-of-sample period is set to 3-months, GMV+COV2 is in the top 5 in all markets except FTSE. However, the observations change slightly when examining 6-month periods. GMV+COV2 drops to 6th place over NIFTY and TOPIX, but stays in the top 5 over DJ, S\&P, and RUSSELL. For the 1-year out-of-sample period, GMV+COV2, on the other hand, jumps into the top 2 spots among the markets DJ, NIFTY, S\&P, and RUSSELL, whereas it disappears entirely from TOPIX. This suggests that GMV+COV2 consistently ranked within the top 5 or 6 across most markets, regardless of variations in out-of-sample periods.  

The model QIS+BS comes next. It is in the top 10 across the 5 markets (DJ, FTSE, S\&P, Russell, and TOPIX) except for NIFTY, for the 3-month period. However, as the investment time increases, its performance decreases: 6-months (in the top 10 over FTSE, S\&P, RUSSELL, and TOPIX) and 1-year (in the top 10 in NIFTY, FTSE, and RUSSELL).

Notably, the classical method of solving the MV model using sample mean and sample covariance fails to find a place in the top 10 for any market in any type of out-of-sample period. The GMV model with sample covariance, on the other hand, does appear in the top 10 for certain cases, such as over TOPIX (for all out-of-sample periods) and RUSSELL (over the 1-year out-of-sample period).
\end{itemize}



\begin{table}[ht]
  \centering
  \caption{Group A- The top 10 shrinkage-based MV and GMV portfolios arranged from highest to lowest ranking order as per their super efficiency scores when all the financial matrices are considered in the (DEA) model. Rankings are given for all six data sets. The Panels 1, 2, and 3, respectively, display the ranking when out-of-sample periods are set to 3-months, 6-months and 1-year in the rolling window scheme. Panel 4 describes the market-best and universal-best performer, along with the geometric mean of efficiency scores (GM of Efficiency) corresponding to each market-best performer.} 
  \label{tab1}
\resizebox{\linewidth}{!}{
  \begin{tabular}{lrrrrrr}
    \toprule
    Rank & DJ  & NIFTY & FTSE & S\&P & RUSSELL & TOPIX\\
    \midrule
    \multicolumn{7}{c}{\textbf{Panel 1: Out-of-sample period= 3 months}} \\
    \midrule
1 & GMV+COV2     & COV2+BS      & LIS+SM    & GMV+COV2     & GMV+QIS      & GMV+COVMKT \\
2 & COV2+QUAD    & GMV+COV2     & COV2+SM   & GMV+COV1     & GMV+COV2      & GMV+SCV \\
3 & COV2+BS      & GMV+COVDIAG  & AS+SM & QIS+SM  & QIS+SM   & COVMKT+BS \\
4 & COV2+JS      & COV1+BS      & LIS+BS        & COV1+JS      & QIS+BS       & GMV+QIS \\
5 & COV2+SM  & COV1+SM  & AS+SM & QIS+BOP     & QIS+BOP      & GMV+COV2 \\
6 & GMV+COV1     & COV2+SM  & GIS+SM    & QIS+JS      & QIS+JS       & QIS+JS \\
7 & COV2+BOP     & GMV+COV1     & GIS+BS        & LS+SM    & GMV+COV1      & GMV+COV1 \\
8 & COV1+QUAD    & GMV+LIS      & QIS+BS       & LS+BOP       & QIS+QUAD     & GMV+LS \\
9 & QIS+BS      & COVDIAG+QUAD & AS+BS     & GMV+QIS     & COV1+SM   & QIS+BS \\
10 & GIS+BS      & LIS+SM   & LS+SM     & QIS+BS      & COV1+JS       & COVCOR+SM \\
        \midrule
    \multicolumn{7}{c}{\textbf{Panel 2: Out-of-sample period= 6 months}} \\
    \midrule
1 & COV2+BS       & GMV+LIS       & QIS+QUAD     & GMV+COV2     & GMV+QIS        & GMV+COVMKT \\
2 & COV2+SM   & LIS+BS        & LIS+QUAD      & COV2+BS      & GMV+COV2        & COVMKT+BS \\
3 & COV2+JS       & AS+BS     & LIS+BS        & QIS+BOP     & QIS+SM     & GMV+SCV \\
4 & GMV+COV2      & COVMKT+BS     & GIS+QUAD      & COV2+JS      & QIS+BS         & GMV+QIS \\
5 & COV2+BOP      & COVDIAG+QUAD  & GIS+BS        & COV2+SM  & COVDIAG+SM  & QIS+BS \\
6 & COV1+SM   & GMV+COV2      & QIS+BS       & COV2+BOP     & QIS+BOP        & GMV+COV2 \\
7 & COV2+QUAD     & AS+SM & LIS+SM    & QIS+BS      & COVDIAG+JS      & QIS+JS \\
8 & COV1+QUAD     & LIS+SM    & AS+QUAD   & QIS+JS      & QIS+JS         & GMV+COV1 \\
9 & COV1+JS       & QIS+SM   & GIS+SM    & COV2+QUAD    & COVDIAG+BS      & COVCOR+JS \\
10 & COV1+BS      & COV2+SM   & QIS+SM   & QIS+QUAD    & COVDIAG+JS      & COVCOR+BS \\
       \midrule
    \multicolumn{7}{c}{\textbf{Panel 3: Out-of-sample period= 1 year}} \\
    \midrule
1 & GMV+COV2     & GMV+COV2      & QIS+SM   & GMV+COV2     & QIS+QUAD     & GMV+COV1 \\
2 & COV2+QUAD    & GMV+LIS       & QIS+BS       & COV2+QUAD    & GMV+COV2      & GMV+QIS \\
3 & COV2+BS      & AS+BS     & LIS+SM    & SCV+BOP   & GMV+QIS      & GMV+SCV \\
4 & COV2+SM  & COV2+BS       & AS+BS     & LS+SM    & GMV+SCV    & GMV+COVMKT \\
5 & COV2+JS      & LIS+BS        & GIS+SM    & COV1+JS      & QIS+SM   & GMV+COVDIAG \\
6 & COV2+BOP     & GIS+BS        & GIS+BS        & LS+QUAD      & QIS+JS       & COVDIAG+JS \\
7 & COV1+QUAD    & QIS+BS       & GIS+QUAD      & COV1+SM  & QIS+BS       & GMV+LS \\
8 & COV1+SM  & COV1+SM       & LIS+QUAD      & LS+BS        & QIS+BOP      & COVMKT+BS \\
9 & GMV+COV1     & AS+SM & QIS+QUAD     & COVDIAG+QUAD & GMV+COV1      & GMV+COVCOR \\
10 & QIS+QUAD   & COV2+SM   & LIS+BS        & SCV+BS    & GMV+LS        & COVCOR+JS \\
\midrule
\multicolumn{7}{c}{\textbf{Panel 4:  Best performing model}} \\
    \midrule
Market best & GMV+COV2     & GMV+COV2      & LIS+SM   & GMV+COV2     & GMV+QIS     & GMV+COVMKT \\
(GM of Efficiency)&(1.01043) &(1.00512)& (1.00285)& (1.07442)& (1.03598)& (1.05409)\\
Universal best &  GMV+COV2         & GMV+COV2            &  GMV+COV2         &  GMV+COV2        & GMV+COV2      & GMV+COV2      \\
     \bottomrule
  \end{tabular}}
\end{table}

\subsection{Assessment for Group B }
 Table \ref{tab2} presents the (DEA) model-based rankings for the top 10 portfolios across all the datasets and varying out-of-sample windows. The following findings are noted from Table \ref{tab2}.
\begin{itemize}
\item \textit{DJ:} COV2+SM model is observed as the best-performing model, signifying the importance of the mean function in the modelling for the return-focused investor. Over this dataset, the QUAD and SAMPLE estimators demonstrated dominance as mean estimators, while COV2 and COV1 exhibited superior performance as variance estimators. 
        
\item \textit{NIFTY:} The portfolio from the COV2+SM model is identified as the best-performing shrinkage model, followed by the  COV2+QUAD and SCV+QUAD models. The QUAD and SAMPLE estimators demonstrated dominance as mean estimators, while COV2 exhibited superior performance as a variance estimator. 
        
\item \textit{FTSE:} LIS+SM model is the best-performing shrinkage model. The QUAD and SAMPLE estimators demonstrated dominance as mean estimators, whereas LIS exhibited superior performance as a variance estimator, followed by GIS and QIS. 
     
\item \textit{S}\&\textit{P :}  Model COV2+SM is identified as the best-performing shrinkage-based MV model. While models utilizing BOP-based mean estimators demonstrated strong performance, they fail to achieve the same level of consistency as the COV2+SM model.         
        
\item \textit{RUSSELL and TOPIX:} The GMV+QIS approach has been recognized as the most effective. This portfolio demonstrates consistent performance across both groups A and B in these two markets. In the RUSSELL market, the QIS+QUAD model yields the second-best portfolio, and the GMV+SCV and GMV+COVMKT models provide the second-best options for TOPIX. 
Notably, in the RUSSELL dataset, the selection of mean estimators has a negligible effect on performance; however, the QIS-based variance estimator showed enhanced efficacy.

\item  \textit{Observations on varying out-of-sample period:} With the exception of TOPIX and RUSSELL (1-year out-of-sample period), the COV2+SM model consistently demonstrates in the top 10 performance across all markets. Based on this output, COV2+SM is the best overall choice, i.~e, a universal best performer, but it fails to be an ideal pick for the market TOPIX. 

The COV2+QUAD follows next. It has been in the top 10 for DJ, NIFTY, RUSSELL, and TOPIX for 3-months, and for DJ, NIFTY, FTSE, and SP for 6-months and 1-year. Furthermore, regardless of the time horizon in low-dimensional markets, the impact of the mean estimator (SAMPLE and QUAD) is more pronounced than that of the covariance estimator; however, in high-dimensional markets, the impact of the covariance estimator is more significant than that of the mean estimator. 

The classical (MV) model with sample estimates, SCV+SM, is a poor performer, with only presence in the top 10 in NIFTY and S\&P (3-month and 1-year out-of-sample periods) and FTSE (3-month out-of-sample period). The classical (GMV) model performs better in the TOPIX, but its performance is poor in other markets. 
\end{itemize}

\begin{table}[htbp]
  \centering
  \caption{Group B- The top 10 shrinkage-based MV and GMV portfolios arranged from highest to lowest ranking order as per their super efficiency scores when only return and risk-adjusted metrics are considered in the (DEA) model. Rankings are given for all six data sets. The Panels 1, 2, and 3, respectively, display the ranking when out-of-sample periods are set to 3-months, 6-months and 1-year in the rolling window scheme. Panel 4 describes the market-best and universal-best performer, along with the geometric mean of efficiency scores (GM of Efficiency) corresponding to each market-best performer.}
  \label{tab2}
\resizebox{\linewidth}{!}{
  \begin{tabular}{lrrrrrr}
    \toprule
    Rank & DJ  & NIFTY & FTSE & S\&P & RUSSELL & TOPIX\\
    \midrule
    \multicolumn{7}{c}{\textbf{Panel 1: Out-of-sample period= 3 months}} \\
    \midrule
1  & COV2+QUAD      & COV2+SM      & LIS+SM    & QIS+BOP      & GMV+QIS     & COVCOR+QUAD \\
2  & COV2+SM    & SCV+SM    & COV2+SM   & LS+BOP        & QIS+QUAD    & GMV+QIS \\
3  & COV1+QUAD      & SCV+QUAD      & GIS+SM    & QIS+SM   & COV2+QUAD    & QIS+JS \\
4  & LIS+QUAD       & COVDIAG+QUAD     & AS+SM & SCV+BOP    & QIS+SM  & GMV+SCV \\
5  & COV1+SM    & COV2+QUAD        & QIS+SM   & QIS+JS       & QIS+BS      & COV2+QUAD \\
6  & GIS+QUAD       & COVDIAG+SM   & LS+SM     & COV2+SM      & QIS+BOP     & QIS+BS \\
7  & QIS+QUAD      & LS+QUAD          & COV1+SM   & LS+SM     & QIS+JS      & QIS+BOP \\
8  & AS+QUAD    & LS+SM        & SCV+SM & QIS+BS       & COV2+BS      & QIS+SM \\
9  & COV2+BS        & COV1+SM      & COVDIAG+SM & SCV+SM & COV2+SM  & QIS+QUAD \\
10 & LIS+SM     & COV1+QUAD        & LIS+QUAD      & COV1+BOP      & COV1+QUAD    & GMV+COVMKT \\

        \midrule
    \multicolumn{7}{c}{\textbf{Panel 2: Out-of-sample period= 6 months}} \\
    \midrule
1  & COV2+SM    & COVDIAG+QUAD   & LIS+QUAD      & COV2+SM    & QIS+SM  & QIS+QUAD \\
2  & COV2+QUAD      & LIS+SM     & QIS+QUAD     & COV2+BOP       & QIS+JS      & GMV+QIS \\
3  & COV1+SM    & COV2+SM    & GIS+QUAD      & COV2+QUAD      & QIS+BS      & GMV+SCV \\
4  & COV1+QUAD      & COV2+QUAD      & COV2+QUAD     & COV2+BS        & QIS+BOP     & QIS+BS \\
5  & COV2+BS        & GIS+QUAD       & AS+QUAD   & QIS+BOP       & COV2+SM  & QIS+JS \\
6  & COVMKT+SM  & GIS+SM     & LIS+SM    & COV2+JS        & COV2+JS      & QIS+BOP \\
7  & LIS+SM     & LIS+QUAD       & GIS+SM    & LS+BOP         & COV2+BS      & QIS+SM \\
8  & QIS+SM    & COVDIAG+SM & AS+QUAD   & COV1+BOP       & COV2+BOP     & GMV+COVMKT \\
9  & GIS+SM     & SCV+QUAD    & COV2+SM   & QIS+SM    & GMV+QIS     & SCV+QUAD \\
10 & COVMKT+QUAD    & AS+SM  & AS+SM & QIS+QUAD      & QIS+QUAD    & COVCOR+QUAD \\
       \midrule

    \multicolumn{7}{c}{\textbf{Panel 3: Out-of-sample period= 1 year}} \\
    \midrule
1  & COV2+QUAD      & COV2+SM    & LIS+SM     & COV2+QUAD     & QIS+QUAD     & GMV+QIS \\
2  & COV2+SM    & AS+SM & GIS+SM     & SCV+BOP    & GMV+QIS      & GMV+SCV \\
3  & COV1+QUAD      & AS+QUAD   & QIS+SM    & COV2+BOP      & GMV+SCV    & QIS+JS \\
4  & COV1+SM    & COV2+QUAD     & GIS+QUAD       & SCV+SM & GMV+COV2      & GMV+COV1 \\
5  & LIS+QUAD       & SCV+SM & LIS+QUAD       & SCV+QUAD   & QIS+SM   & QIS+BS \\
6  & QIS+QUAD      & SCV+QUAD   & QIS+QUAD      & COV2+SM   & QIS+JS       & GMV+LS \\
7  & GIS+QUAD       & COV1+SM   & AS+SM  & COV2+BS       & QIS+BS       & GMV+COVMKT \\
8  & QIS+SM    & COV1+QUAD     & AS+QUAD    & SCV+QUAD   & QIS+BOP      & QIS+SM \\
9  & LIS+SM     & QIS+QUAD     & COV2+SM    & SCV+JS     & GMV+COV1      & QIS+BOP \\
10 & GIS+SM     & QIS+SM   & COV2+QUAD      & LS+BOP        & SCV+QUAD   & SCV+JS \\
\midrule

\multicolumn{7}{c}{\textbf{Panel 4: Best performing model}} \\
    \midrule
Market best & COV2+SM     & COV2+SM      & LIS+SM   & COV2+SM     & GMV+QIS     & GMV+QIS \\
(GM of Efficiency)&(1.00165) &(1.00077)& (1.00241)& (1.01038)& (1.00473)& (1.01921)\\
Universal best &  COV2+SM         & COV2+SM          &  COV2+SM         &  COV2+SM        & COV2+SM      & COV2+SM      \\
\bottomrule
     \end{tabular}}
\end{table}

\subsection{Assessment for Group C}
Table \ref{tab3} organises the efficiency scores for the top 10 portfolios across all datasets and varying out-of-sample windows. 
Table \ref{tab3} concludes the following results. 
\begin{itemize}
    \item \textit{DJ, NIFTY and FTSE:} In DJ, the GMV+COV2 model shows consistent best performance, and among the (MV) models,  the COV2+BOP achieves the highest performance. Across the NIFTY and FTSE datasets, portfolios from the GMV+LIS model perform best, followed by GMV+COVCOR, which is second best in NIFTY and GMV+GIS in FTSE. Notably, portfolios from the GMV models consistently outperformed others in risk mitigation, as evidenced by the presence of at least three GMV models within the top 10 ranks. In contrast, models based on sample mean and variance estimators exhibited poor performance.

    \item \textit{S}\&\textit{P, RUSSELL, and TOPIX:} Portfolio from the GMV+COV2 model is identified as the best-performing model over the three high-dimensional data sets, S\&P, RUSSELL, and TOPIX. The second best choice is GMV+QIS over the data sets, S\&P, and RUSSELL, whereas GMV+COVCOR is the second choice for TOPIX. Similar to the set of low-dimensional markets, here also portfolios from the GMV-based model consistently outperformed others in risk mitigation, as evidenced by the presence of at least six GMV models within the top 10 rankings across all periods. In contrast to low-dimensional datasets performance of GMV+SCV is quite impressive.

    \item \textit{Varying out-of-sample periods:} Portfolios from the model GMV+COV2 always show up in the top 10, except for a few times, like FTSE over 3-months and NIFTY over 1-year out-of-sample period. It actually ranks in the top three for the DJ, S\&P, RUSSELL, and TOPIX, regardless of the length of the investment period. In light of this, GMV+COV2 becomes a universal-best performer.

The next contender is GMV+QIS, which features very often in the top 10 for several markets (except to FTSE over 6-months and DJ, NIFTY \& TOPIX over a 1-year out-of-sample period). Meanwhile, GMV+COVMKT is in the top 10 in all markets for all out-of-sample periods except for FTSE and NIFTY for one year. The SAMPLE estimate for covariance in the GMV framework i.~e. GMV+SCV occurs exclusively in high-dimensional datasets.
        
Considering group C primarily targets to risk-averse investors, GMV-based models are at the top of the rankings in almost all markets, especially FTSE, S\&P, RUSSELL, and TOPIX.

    \end{itemize}

\begin{table}[htbp]
  \centering
  \caption{Group C- The top 10 shrinkage-based MV and GMV portfolios arranged from highest to lowest ranking order as per their super efficiency scores when only risk metrics are considered in the (DEA) model. Rankings are given for all six data sets. The Panels 1, 2, and 3, respectively, display the ranking when out-of-sample periods are set to 3-months, 6-months and 1-year in the rolling window scheme. Panel 4 describes the market-best and universal-best performer, along with the geometric mean of efficiency scores (GM of Efficiency) corresponding to each market-best performer.}
  \label{tab3}
\resizebox{\linewidth}{!}{
  \begin{tabular}{lrrrrrr}
    \toprule
    Rank & DJ  & NIFTY & FTSE & S\&P & RUSSELL & TOPIX\\
    \midrule
    \multicolumn{7}{c}{\textbf{Panel 1: Out-of-sample period= 3 months}} \\
    \midrule
1  & GMV+COV2     & COVDIAG+BOP   & QIS+BS       & GMV+COV2     & GMV+COV2     & GMV+COV2 \\
2  & COV2+BOP     & GMV+LIS       & GMV+LIS       & GMV+QIS     & GMV+QIS     & GMV+COVMKT \\
3  & GMV+COVMKT   & GMV+COVMKT    & GMV+COVCOR    & GMV+COV1     & GMV+SCV   & GMV+COVCOR \\
4  & COVMKT+BOP   & COVMKT+BOP    & LIS+BS        & GMV+LS       & GMV+COVMKT   & GMV+SCV \\
5  & COVMKT+JS    & GMV+COVCOR    & GMV+GIS       & GMV+SCV   & GMV+COV1     & GMV+COV1 \\
6  & GMV+AS   & GMV+AS    & COVCOR+JS     & GMV+COVMKT   & GMV+COVDIAG  & GMV+QIS \\
7  & GMV+QIS     & COVCOR+JS     & LIS+JS        & GMV+COVDIAG  & GMV+LS       & COVCOR+JS \\
8  & GMV+GIS      & GMV+QIS      & QIS+JS       & QIS+BS      & GMV+COVCOR   & GMV+LS \\
9  & GMV+LIS      & GMV+GIS       & GIS+JS        & QIS+JS      & COVMKT+SM & GMV+COVDIAG \\
10 & LIS+BOP      & GMV+COV2      & GMV+QIS      & QIS+SM  & COVMKT+JS    & COVCOR+BS \\

        \midrule
    \multicolumn{7}{c}{\textbf{Panel 2: Out-of-sample period= 6 months}} \\
    \midrule
1  & GMV+COV2     & GMV+COVMKT    & COVCOR+BS     & GMV+COV2     & GMV+COV1     & GMV+COV2 \\
2  & COVDIAG+BOP  & GMV+LIS       & GMV+COV2      & GMV+QIS     & GMV+QIS     & GMV+COVCOR \\
3  & GMV+COVMKT   & GMV+COVCOR    & GMV+LIS       & GMV+COV1     & GMV+COV2     & GMV+COVMKT \\
4  & COVDIAG+JS   & GMV+AS    & GMV+GIS       & QIS+JS      & GMV+SCV   & GMV+COV1 \\
5  & GMV+AS   & COVCOR+JS     & GIS+JS        & QIS+BS      & GMV+COVMKT   & GMV+SCV \\
6  & COV2+JS      & GMV+GIS       & QIS+JS       & GMV+COVMKT   & GMV+LS       & GMV+LS \\
7  & QIS+BOP     & GMV+COV2      & LIS+JS        & GMV+SCV   & GMV+COVDIAG  & GMV+LS \\
8  & COVMKT+BOP   & GMV+QIS      & LIS+JS        & QIS+SM  & GMV+COVCOR   & COVCOR+BS \\
9  & COV2+BOP     & COVMKT+JS     & COVMKT+JS     & GMV+LS       & COVDIAG+SM & GMV+COVDIAG \\
10 & GMV+QIS     & LIS+JS        & GMV+AS    & QIS+QUAD    & COVDIAG+JS   & GMV+QIS \\
       \midrule

    \multicolumn{7}{c}{\textbf{Panel 3: Out-of-sample period= 1 year}} \\
    \midrule
1  & COV2+BOP     & GMV+LIS        & GMV+COV2      & GMV+COV2     & GMV+SCV   & GMV+COV2 \\
2  & COVMKT+BOP   & LIS+BOP        & GMV+GIS       & GMV+QIS     & GMV+COV2     & GMV+COV1 \\
3  & GMV+COV2     & COVMKT+BOP     & GMV+LIS       & GMV+COVCOR   & GMV+QIS     & GMV+COVCOR \\
4  & GMV+COVMKT   & AS+BOP     & GMV+QIS      & GMV+COV1     & GMV+COVMKT   & GMV+SCV \\
5  & COVMKT+JS    & QIS+BOP       & GMV+AS    & GMV+COVMKT   & GMV+COV1     & GMV+COVMKT \\
6  & COV2+JS      & SCV+BOP     & GMV+COV1      & GMV+COVDIAG  & GMV+LS       & GMV+LS \\
7  & GMV+COVCOR   & GIS+BOP        & LIS+JS        & GMV+SCV   & QIS+SM  & COVCOR+JS \\
8  & AS+BOP   & COVDIAG+BOP    & GMV+COVMKT    & QIS+JS      & QIS+JS      & COVCOR+BS \\
9  & QIS+BOP     & GMV+COVCOR     & GIS+JS        & COVMKT+BS    & QIS+BS      & GMV+QIS \\
10 & GIS+BOP      & GMV+GIS        & GMV+COVCOR    & LS+BS        & QIS+BOP     & GMV+COVDIAG \\
\midrule

\multicolumn{7}{c}{\textbf{Panel 4: Best performing model}} \\
    \midrule
Market best & GMV+COV2     &  GMV+LIS    &  GMV+LIS  &  GMV+COV2   &  GMV+COV2     &  GMV+COV2 \\
(GM of Efficiency)&(1.00913) &(1.00355)& (1.00149)& (1.06506)& (1.01091)& (1.09658)\\
Universal best &   GMV+COV2         & GMV+COV2        &   GMV+COV2      &  GMV+COV2   & GMV+COV2      &  GMV+COV2  \\

     \bottomrule
  \end{tabular}}
\end{table}

\section{Comparison with Benchmark Models}\label{sec7}

Tables \ref{tab4}-\ref{tab6} report the financial performance of the five optimal portfolios from benchmark PO models, the shrinkage-based market best performer, and the universal best performer across all the datasets. The tables also present super-efficiency scores of portfolios from groups A, B, and C, under the headings Efficiency-A, Efficiency-B, and Efficiency-C, respectively. The following conclusions are drawn from the tables.

\begin{itemize}
\item \textit{DJ:} The GMV+COV2 model, which is identified as both the market and universal best-performing method over the data set DJ for the groups A and C, demonstrates superior efficiency relative to all the benchmark PO models in all three out-of-sample periods. However, for group B, the COV2+SM model, which is selected as the top performer for DJ as well as the universal best performer, outperforms all except MM over the 3-month out-of-sample period. 

\item \textit{NIFTY:} The GMV+COV2, which is the market-best as well as the universal best-performing model for group A, demonstrates superior efficiency relative to benchmark models except for classical MV. For group B, the COV2+SM model, the market best and the universal best performer, outperforms all benchmark models in all out-of-sample periods. For group C, the market-best performer, GMV+LIS demonstrates superior efficiency relative to all benchmark models in all out-of-sample periods.  The universal best-performing GMV+COV2 for group C is seen to outperform all benchmark models for the 6-month out-of-sample period. However, for the 3-month and 1-year out-of-sample periods, classical GMV has better super-efficiency.

\item \textit{FTSE:} The market best performer for groups A and B over FTSE, the LIS+SM model performs better except for the CVaR model in group A for a 3-month out-of-sample period and MM over the 6-month out-of-sample period. LIS+SM exceeds the performance of classical MV, GMV, and SMAD in the 3-month and 6-month out-of-sample periods while surpassing all benchmarks in the 1-year out-of-sample period. The market best performer for group C,  GMV+LIS, emerged as the superior model over all benchmark models in all out-of-sample periods.
    
The GMV+COV2 model performs poorly with this data set.  The FTSE market acted differently from other markets, ignoring GMV+COV2 for groups A and B altogether. It only chooses it for group A over the 1-year out-of-sample period. In fact, GMV+COV2 performs better than other benchmark models in group C. For group B, the universal best performer, COV2+SM, outperforms classical MV, GMV, and SMAD in both the 3-month and 6-month out-of-sample periods, and it also outperforms CVaR in the 1-year out-of-sample period.

\item \textit{S}\&\textit{P:} For both groups A and C, the GMV+COV2 model,  the best-performing model (market as well as universal), demonstrates a superior efficiency score relative to all the other benchmark models except for MM in group A for the case of the 1-year out-of-sample period. This indicates that the portfolio from the model GMV+COV2 is a robust choice for both groups A and C in this market. 
For Group B, the COV2+SM model (market best as well as the universal best performer) has a superior efficiency score relative to all the other benchmark models, except for classical MV in the 3-month out-of-sample period and MM in the 3-month and 1-year out-of-sample periods. 

\item \textit{RUSSELL:} For both groups A and B, the GMV+QIS model demonstrates its superior efficiency relative to all the other benchmark models except for MM for group A in the 3-month and 6-month out-of-sample periods and except for CVaR for the 1-year period. For Group B, GMV+QIS outperforms the benchmark models in all out-of-sample periods. On the other hand, for group C, the GMV+COV2 model (market best as well as the universal best performer) surpasses all the benchmark models except SMAD in all periods.

The GMV+COV2, which is selected as the universal best-performer for group A, demonstrates its superior efficiency relative to all the other benchmark models in all 3 out-of-sample periods except for MM in the 3-month and 6-month out-of-sample periods. For group B, the universal best-performing model, COV2+SM, outperforms all benchmark models in the 3-month and 6-month out-of-sample periods. However, in the 1-year out-of-sample period, it outperforms only SMAD.

\item \textit{TOPIX:} GMV+COVMKT  (market best performer) model for group A is superior in efficiency relative to all benchmark models except SMAD in all periods. For Group B, GMV+QIS (the market's best performer) outperforms all benchmark models in the 1-year out-of-sample period; however, in the 3-month and 6-month out-of-sample periods, MM surpasses it. The COV2+SM (universal best performer) surpasses only the CVaR portfolio in all periods. For group C, the market and universal best-performing model, GMV+COV2, exhibits superior efficiency than all benchmark models, except SMAD, in all periods. Additionally, classical GMV also performs well in the 3-month out-of-sample period. GMV+COV2, which is universally best for group A, demonstrates superior efficiency relative to CVaR and MM in all out-of-sample periods.
\end{itemize}

\begin{center}
\captionsetup{width=\textwidth}
{\tiny\begin{longtable}{|c|ccccc|cccccc|}
\caption{Comparative study with benchmark models: Out-of-sample performance analysis of the selected market-best and universal-best performer with the five benchmark models when the out-of-sample period is set to 3-months. The comparison analysis is based on the efficiency scores from the (DEA) model with respect to all the groups of investors, A, B, and C. The analysis is presented for all six data sets. }\label{tab4} \\ \hline
\multirow{3}{*}{\textbf{Efficiency}}
& \multicolumn{5}{c|}{\textbf{Benchmark Models}}
& \multicolumn{6}{c|}{\textbf{Shrinkage-based}} \\
\cline{2-12}
& MV & CVaR & SMAD & MM & GMV
& \multicolumn{2}{c|}{\textbf{Group A}}
& \multicolumn{2}{c|}{\textbf{Group B}}
& \multicolumn{2}{c|}{\textbf{Group C}} \\
\cline{7-12}
& & &  &  &
& Market & Universal & Market & Universal & Market & Universal \\
\hline
\endfirsthead

\caption[]{(continued)}\\
\hline
\multirow{3}{*}{\shortstack{\textbf{Performance} \\ \textbf{Measure}}}
& \multicolumn{5}{c|}{\textbf{Benchmark Models}}
& \multicolumn{6}{c|}{\textbf{Shrinkage-based}} \\
\cline{2-12}
& MV & CVaR & SMAD & MIN & GMV+
& \multicolumn{2}{c|}{\textbf{Group A}}
& \multicolumn{2}{c|}{\textbf{Group B}}
& \multicolumn{2}{c|}{\textbf{Group C}} \\
\cline{7-12}
& & &  & MAX &SCV
& Market & Universal & Market & Universal & Market & Universal \\
\hline
\endhead

\multicolumn{6}{|c|}{\multirow{2}{*}{\textbf{DJ}}} &
\textbf{GMV+} & \textbf{GMV+} & \textbf{COV2+} & \textbf{COV2+} & \textbf{GMV+} & \textbf{GMV+} \\
\multicolumn{6}{|c|}{} &
\textbf{COV2} & \textbf{COV2} & \textbf{SM}   & \textbf{SM}   & \textbf{COV2} & \textbf{COV2} \\
\hline
Efficiency-A & 0.90130 & 0.84430 & 0.89500 & 0.95960 & 0.87950 & \textbf{1.27690} & \textbf{1.27690} & 1.04150 & 1.04150 & \textbf{1.27690} & \textbf{1.27690} \\
Efficiency-B & 0.88480 & 0.83390 & 0.85790 & \textbf{1.09390} & 0.82180 & 0.92950 & 0.92950 & 1.02690 & 1.02690 & 0.92950 & 0.92950 \\
Efficiency-C & 0.97210 & 0.96010 & 0.98260 & 0.83410 & 1.00420 & \textbf{1.17520} & \textbf{1.17520} & 0.96180 & 0.96180 & \textbf{1.17520} & \textbf{1.17520} \\
\hline
\multicolumn{6}{|c|}{\multirow{2}{*}{\textbf{NIFTY}}} &
\textbf{GMV+} & \textbf{GMV+} & \textbf{COV2+} & \textbf{COV2+} & \textbf{GMV+} & \textbf{GMV+} \\
\multicolumn{6}{|c|}{} &
\textbf{COV2} & \textbf{COV2} & \textbf{SM}   & \textbf{SM}   & \textbf{LIS} & \textbf{COV2} \\
\hline
Efficiency-A & 1.01870 & 0.80790 & 0.99690 & 0.67870 & 1.00790 & 1.01110 & 1.01110 & \textbf{1.02650} & \textbf{1.02650} & 1.00140 & 1.01110 \\
Efficiency-B  & 1.00170 & 0.86820 & 0.94960 & 0.77470 & 0.95200 & 0.94170 & 0.94170 & \textbf{1.01510} & \textbf{1.01510} & 0.93480 & 0.94170 \\
Efficiency-C    & 0.96110 & 0.88080 & 0.99720 & 0.82930 & 1.00000 & 0.99980 & 0.99980 & 0.94610 & 0.94610 & \textbf{1.00690} & 0.99980 \\
\hline
\multicolumn{6}{|c|}{\multirow{2}{*}{\textbf{FTSE}}} &
\textbf{LIS+} & \textbf{GMV+} & \textbf{LIS+} & \textbf{COV2+} & \textbf{GMV+} & \textbf{GMV+} \\
\multicolumn{6}{|c|}{} &
\textbf{SM} & \textbf{COV2} & \textbf{SM}   & \textbf{SM}   & \textbf{LIS} & \textbf{COV2} \\
\hline
Efficiency-A   & 0.96130 & \textbf{1.18540} & 0.68650 & 0.91260 & 0.78420 & 1.03530 & 0.89170 & 1.03530 & 1.00030 & 0.93070 & 0.89170 \\
Efficiency-B   & 0.86940 & \textbf{1.17000} & 0.60550 & 1.02110 & 0.68320 & 0.90520 & 0.69100 & 0.90520 & 0.89530 & 0.71180 & 0.69100 \\
Efficiency-C   & 0.98410 & 0.96050 & \textbf{1.01660} & 0.82180 & 0.99440 & 0.97930 & 0.99660 & 0.97930 & 0.98120 & 1.00910 & 0.99660 \\
\hline
\multicolumn{6}{|c|}{\multirow{2}{*}{\textbf{S\&P}}} &
\textbf{GMV+} & \textbf{GMV+} & \textbf{COV2+} & \textbf{COV2+} & \textbf{GMV+} & \textbf{GMV+} \\
\multicolumn{6}{|c|}{} &
\textbf{COV2} & \textbf{COV2} & \textbf{SM}   & \textbf{SM}   & \textbf{COV2} & \textbf{COV2} \\
\hline
Efficiency-A  & 1.04850 & 0.97660 & 0.93650 & 1.00310 & 0.97330 & \textbf{1.21110} & \textbf{1.21110} & 1.00940 & 1.00940 & \textbf{1.21110} & \textbf{1.21110} \\
Efficiency-B  & 1.01920 & 0.97670 & 0.82180 & \textbf{1.06130} & 0.92560 & 0.94290 & 0.94290 & 0.99180 & 0.99180 & 0.94290 & 0.94290 \\
Efficiency-C  & 0.96860 & 0.90110 & 1.10460 & 0.88390 & 0.99660 & \textbf{1.17890} & \textbf{1.17890} & 0.96740 & 0.96740 & \textbf{1.17890} & \textbf{1.17890} \\
\hline
\multicolumn{6}{|c|}{\multirow{2}{*}{\textbf{RUSSELL}}} &
\textbf{GMV+} & \textbf{GMV+} & \textbf{GMV+} & \textbf{COV2+} & \textbf{GMV+} & \textbf{GMV+} \\
\multicolumn{6}{|c|}{} &
\textbf{QIS} & \textbf{COV2} & \textbf{QIS}   & \textbf{SM}   & \textbf{COV2} & \textbf{COV2} \\
\hline
Efficiency-A & 0.94610 & 0.94460 & 0.67640 & 1.04260 & 0.92680 & \textbf{1.05360} & 1.02220 & \textbf{1.05360} & 0.95180 & 1.02220 & 1.02220 \\
Efficiency-B & 0.95460 & 0.86830 & 0.55350 & 0.96210 & 0.92160 & \textbf{1.03980} & 0.98550 & \textbf{1.03980} & 0.97850 & 0.98550 & 0.98550 \\
Efficiency-C & 0.97210 & 1.03180 & \textbf{1.11860} & 0.98260 & 0.99490 & 1.01150 & 1.03720 & 1.01150 & 0.96000 & 1.03720 & 1.03720 \\
\hline
\multicolumn{6}{|c|}{\multirow{2}{*}{\textbf{TOPIX}}} &
\textbf{GMV+} & \textbf{GMV+} & \textbf{GMV+} & \textbf{COV2+} & \textbf{GMV+} & \textbf{GMV+} \\
\multicolumn{6}{|c|}{} &
\textbf{COVMKT} & \textbf{COV2} & \textbf{QIS}   & \textbf{SM}   & \textbf{COV2} & \textbf{COV2} \\
\hline
Efficiency-A & 0.95760 & 0.78100 & \textbf{1.20080} & 0.90060 & 1.01880 & 1.10590 & 0.93390 & 1.01740 & 0.81200 & 0.93390 & 0.93390 \\
Efficiency-B & 0.99770 & 0.87150 & 0.89410 & \textbf{1.10420} & 1.00690 & 0.99820 & 0.87830 & 1.01720 & 0.91810 & 0.87830 & 0.87830 \\
Efficiency-C & 0.88060 & 0.73260 & \textbf{1.12630} & 0.73710 & 1.01180 & 1.05200 & 1.00520 & 0.99590 & 0.71280 & 1.00520 & 1.00520 \\
\hline

\end{longtable}}
\end{center}

\begin{center}
\captionsetup{width=\textwidth}
{\tiny\begin{longtable}{|c|ccccc|cccccc|}
\caption{ Comparative study with benchmark models: Out-of-sample performance analysis of the selected market-best and universal-best performer with the five benchmark models when the out-of-sample period is set to 6-months. The comparison analysis is based on the efficiency scores from the (DEA) model with respect to all the groups of investors, A, B, and C. The analysis is presented for all six data sets.}\label{tab5} \\ \hline
\multirow{3}{*}{\textbf{Efficiency}}
& \multicolumn{5}{c|}{\textbf{Benchmark Models}}
& \multicolumn{6}{c|}{\textbf{Shrinkage-based}} \\
\cline{2-12}
& MV & CVaR & SMAD & MM & GMV
& \multicolumn{2}{c|}{\textbf{Group A}}
& \multicolumn{2}{c|}{\textbf{Group B}}
& \multicolumn{2}{c|}{\textbf{Group C}} \\
\cline{7-12}
& & &  &  &
& Market & Universal & Market & Universal & Market & Universal \\
\hline
\endfirsthead

\caption[]{(continued)}\\
\hline
\multirow{3}{*}{\shortstack{\textbf{Performance} \\ \textbf{Measure}}}
& \multicolumn{5}{c|}{\textbf{Benchmark Models}}
& \multicolumn{6}{c|}{\textbf{Shrinkage-based}} \\
\cline{2-12}
& MV & CVaR & SMAD & MIN & GMV+
& \multicolumn{2}{c|}{\textbf{Group A}}
& \multicolumn{2}{c|}{\textbf{Group B}}
& \multicolumn{2}{c|}{\textbf{Group C}} \\
\cline{7-12}
& & &  & MAX &SCV
& Market & Universal & Market & Universal & Market & Universal \\
\hline
\endhead

\multicolumn{6}{|c|}{\multirow{2}{*}{\textbf{DJ}}} &
\textbf{GMV+} & \textbf{GMV+} & \textbf{COV2+} & \textbf{COV2+} & \textbf{GMV+} & \textbf{GMV+} \\
\multicolumn{6}{|c|}{} &
\textbf{COV2} & \textbf{COV2} & \textbf{SM}   & \textbf{SM}   & \textbf{COV2} & \textbf{COV2} \\
\hline
Efficiency-A & 0.93380 & 0.86060 & 0.85850 & 0.83560 & 0.85890 & \textbf{1.15350} & \textbf{1.15350} & 1.02170 & 1.02170 & \textbf{1.15350} & \textbf{1.15350} \\
Efficiency-B & 0.90670 & 0.84710 & 0.82540 & 0.96550 & 0.80710 & 0.91440 & 0.91440 & \textbf{1.02110} & \textbf{1.02110} & 0.91440 & 0.91440 \\
Efficiency-C & 0.96820 & 0.96730 & 0.97630 & 0.82170 & 1.00070 & \textbf{1.16950} & \textbf{1.16950} & 0.96040 & 0.96040 & \textbf{1.16950} & \textbf{1.16950} \\
\hline
\multicolumn{6}{|c|}{\multirow{2}{*}{\textbf{NIFTY}}} &
\textbf{GMV+} & \textbf{GMV+} & \textbf{COV2+} & \textbf{COV2+} & \textbf{GMV+} & \textbf{GMV+} \\
\multicolumn{6}{|c|}{} &
\textbf{COV2} & \textbf{COV2} & \textbf{SM}   & \textbf{SM}   & \textbf{LIS} & \textbf{COV2} \\
\hline
Efficiency-A & 1.02700 & 0.82510 & 0.96070 & 0.72300 & 1.00020 & 1.01180 & 1.01180 & \textbf{1.04140} & \textbf{1.04140} & 1.00590 & 1.01180 \\
Efficiency-B & 1.00270 & 0.88190 & 0.90930 & 0.80850 & 0.93720 & 0.93570 & 0.93570 & \textbf{1.02150} & \textbf{1.02150} & 0.93040 & 0.93570 \\
Efficiency-C & 0.95750 & 0.87820 & 0.98570 & 0.83690 & 1.00080 & 1.00300 & 1.00300 & 0.94480 & 0.94480 & \textbf{1.01160} & 1.00300 \\
\hline
\multicolumn{6}{|c|}{\multirow{2}{*}{\textbf{FTSE}}} &
\textbf{LIS+} & \textbf{GMV+} & \textbf{LIS+} & \textbf{COV2+} & \textbf{GMV+} & \textbf{GMV+} \\
\multicolumn{6}{|c|}{} &
\textbf{SM} & \textbf{COV2} & \textbf{SM}   & \textbf{SM}   & \textbf{LIS} & \textbf{COV2} \\
\hline
Efficiency-A & 0.94450 & 0.99260 & 0.72370 & \textbf{1.17560} & 0.82200 & 1.03310 & 0.93040 & 1.03310 & 1.01110 & 0.95060 & 0.93040 \\
Efficiency-B & 0.86940 & 0.94440 & 0.64770 & \textbf{1.32360} & 0.75190 & 0.94190 & 0.76160 & 0.94190 & 0.93290 & 0.78270 & 0.76160 \\
Efficiency-C & 0.97820 & 0.93350 & 1.00330 & 0.82450 & 0.99390 & 0.96740 & \textbf{1.00600} & 0.96740 & 0.97110 & 1.00590 & \textbf{1.00600} \\
\hline
\multicolumn{6}{|c|}{\multirow{2}{*}{\textbf{S\&P}}} &
\textbf{GMV+} & \textbf{GMV+} & \textbf{COV2+} & \textbf{COV2+} & \textbf{GMV+} & \textbf{GMV+} \\
\multicolumn{6}{|c|}{} &
\textbf{COV2} & \textbf{COV2} & \textbf{SM}   & \textbf{SM}   & \textbf{COV2} & \textbf{COV2} \\
\hline
Efficiency-A & 0.97100 & 0.80400 & 0.83970 & 0.92030 & 0.86660 & \textbf{1.12720} & \textbf{1.12720} & 1.02030 & 1.02030 & \textbf{1.12720} & \textbf{1.12720} \\
Efficiency-B & 0.95600 & 0.77760 & 0.75420 & 0.92340 & 0.83330 & 0.90930 & 0.90930 & \textbf{1.00760} & \textbf{1.00760} & 0.90930 & 0.90930 \\
Efficiency-C & 0.98040 & 0.93410 & 1.07000 & 0.91700 & 0.99650 & \textbf{1.23600} & \textbf{1.23600} & 0.97680 & 0.97680 & \textbf{1.23600} & \textbf{1.23600} \\
\hline
\multicolumn{6}{|c|}{\multirow{2}{*}{\textbf{RUSSELL}}} &
\textbf{GMV+} & \textbf{GMV+} & \textbf{GMV+} & \textbf{COV2+} & \textbf{GMV+} & \textbf{GMV+} \\
\multicolumn{6}{|c|}{} &
\textbf{QIS} & \textbf{COV2} & \textbf{QIS}   & \textbf{SM}   & \textbf{COV2} & \textbf{COV2} \\
\hline
Efficiency-A & 0.97680 & 0.96500 & 0.90880 & \textbf{1.06600} & 0.92960 & 1.05050 & 1.00530 & 1.05050 & 0.98350 & 1.00530 & 1.00530 \\
Efficiency-B & 0.98450 & 0.93570 & 0.80340 & 0.98830 & 0.92550 & \textbf{1.03020} & 0.97900 & 1.03020 & 1.01430 & 0.97900 & 0.97900 \\
Efficiency-C & 0.96500 & 0.95940 & \textbf{1.12200} & 0.96900 & 0.99190 & 1.01110 & 1.02690 & 1.01110 & 0.95800 & 1.02690 & 1.02690 \\
\hline
\multicolumn{6}{|c|}{\multirow{2}{*}{\textbf{TOPIX}}} &
\textbf{GMV+} & \textbf{GMV+} & \textbf{GMV+} & \textbf{COV2+} & \textbf{GMV+} & \textbf{GMV+} \\
\multicolumn{6}{|c|}{} &
\textbf{COVMKT} & \textbf{COV2} & \textbf{QIS}   & \textbf{SM}   & \textbf{COV2} & \textbf{COV2} \\
\hline
Efficiency-A & 0.94820 & 0.78150 & \textbf{1.27930} & 0.88730 & 1.03010 & 1.07020 & 0.95500 & 1.02020 & 0.79580 & 0.95500 & 0.95500 \\
Efficiency-B & 0.99340 & 0.88540 & 0.95500 & \textbf{1.03170} & 1.00650 & 0.98510 & 0.85830 & 1.02370 & 0.93250 & 0.85830 & 0.85830 \\
Efficiency-C & 0.90300 & 0.80590 & \textbf{1.09630} & 0.82140 & 1.00970 & 1.03740 & 1.05020 & 0.99390 & 0.75480 & 1.05020 & 1.05020 \\
\hline

\end{longtable}}
\end{center}

\begin{center}
\captionsetup{width=\textwidth}
{\tiny\begin{longtable}{|c|ccccc|cccccc|}
\caption{ Comparative study with benchmark models: Out-of-sample performance analysis of the selected market-best and universal-best performer with the five benchmark models when the out-of-sample period is set to 1-year. The comparison analysis is based on the efficiency scores from the (DEA) model with respect to all the groups of investors, A, B, and C. The analysis is presented for all six data sets.}\label{tab6} \\ \hline
\multirow{3}{*}{\textbf{Efficiency}}
& \multicolumn{5}{c|}{\textbf{Benchmark Models}}
& \multicolumn{6}{c|}{\textbf{Shrinkage-based}} \\
\cline{2-12}
& MV & CVaR & SMAD & MM & GMV
& \multicolumn{2}{c|}{\textbf{Group A}}
& \multicolumn{2}{c|}{\textbf{Group B}}
& \multicolumn{2}{c|}{\textbf{Group C}} \\
\cline{7-12}
& & &  &  &
& Market & Universal & Market & Universal & Market & Universal \\
\hline
\endfirsthead

\caption[]{(continued)}\\
\hline
\multirow{3}{*}{\shortstack{\textbf{Performance} \\ \textbf{Measure}}}
& \multicolumn{5}{c|}{\textbf{Benchmark Models}}
& \multicolumn{6}{c|}{\textbf{Shrinkage-based}} \\
\cline{2-12}
& MV & CVaR & SMAD & MIN & GMV+
& \multicolumn{2}{c|}{\textbf{Group A}}
& \multicolumn{2}{c|}{\textbf{Group B}}
& \multicolumn{2}{c|}{\textbf{Group C}} \\
\cline{7-12}
& & &  & MAX &SCV
& Market & Universal & Market & Universal & Market & Universal \\
\hline
\endhead

\multicolumn{6}{|c|}{\multirow{2}{*}{\textbf{DJ}}} &
\textbf{GMV+} & \textbf{GMV+} & \textbf{COV2+} & \textbf{COV2+} & \textbf{GMV+} & \textbf{GMV+} \\
\multicolumn{6}{|c|}{} &
\textbf{COV2} & \textbf{COV2} & \textbf{SM}   & \textbf{SM}   & \textbf{COV2} & \textbf{COV2} \\
\hline
Efficiency-A & 0.89950 & 0.80810 & 0.86700 & 0.75410 & 0.81420 & \textbf{1.10170} & \textbf{1.10170} & 0.99880 & 0.99880 & \textbf{1.10170} & \textbf{1.10170} \\
Efficiency-B & 0.88470 & 0.77960 & 0.81580 & 0.84220 & 0.76660 & 0.89340 & 0.89340 & \textbf{0.99770} & \textbf{0.99770} & 0.89340 & 0.89340 \\
Efficiency-C & 0.97090 & 0.98720 & 0.98500 & 0.86510 & 1.00050 & \textbf{1.14360} & \textbf{1.14360} & 0.96640 & 0.96640 & \textbf{1.14360} & \textbf{1.14360} \\
\hline
\multicolumn{6}{|c|}{\multirow{2}{*}{\textbf{NIFTY}}} &
\textbf{GMV+} & \textbf{GMV+} & \textbf{COV2+} & \textbf{COV2+} & \textbf{GMV+} & \textbf{GMV+} \\
\multicolumn{6}{|c|}{} &
\textbf{COV2} & \textbf{COV2} & \textbf{SM}   & \textbf{SM}   & \textbf{LIS} & \textbf{COV2} \\
\hline
Efficiency-A & 1.02000 & 0.78310 & 0.98540 & 0.68520 & 1.00680 & 1.01090 & 1.01090 & \textbf{1.03370} & \textbf{1.03370} & 1.00320 & 1.01090 \\
Efficiency-B & 1.00160 & 0.84010 & 0.92960 & 0.75150 & 0.93530 & 0.93750 & 0.93750 & \textbf{1.01930} & \textbf{1.01930} & 0.93230 & 0.93750 \\
Efficiency-C & 0.95030 & 0.87610 & 0.99390 & 0.84290 & 1.00520 & 1.00110 & 1.00110 & 0.93990 & 0.93990 & \textbf{1.00780} & 1.00110 \\
\hline
\multicolumn{6}{|c|}{\multirow{2}{*}{\textbf{FTSE}}} &
\textbf{LIS+} & \textbf{GMV+} & \textbf{LIS+} & \textbf{COV2+} & \textbf{GMV+} & \textbf{GMV+} \\
\multicolumn{6}{|c|}{} &
\textbf{SM} & \textbf{COV2} & \textbf{SM}   & \textbf{SM}   & \textbf{LIS} & \textbf{COV2} \\
\hline
Efficiency-A & 0.96600 & 0.65380 & 0.86370 & 0.88880 & 0.96560 & \textbf{1.04740} & 0.98700 & \textbf{1.04740} & 0.98230 & 1.00030 & 0.98700 \\
Efficiency-B & 0.95680 & 0.69400 & 0.82820 & 0.99990 & 0.91650 & \textbf{1.03330} & 0.89620 & \textbf{1.03330} & 0.98160 & 0.91420 & 0.89620 \\
Efficiency-C & 0.95530 & 0.90230 & 0.98980 & 0.85390 & 0.99130 & 0.96090 & \textbf{1.00840} & 0.96090 & 0.95300 & 1.00740 & \textbf{1.00840} \\
\hline
\multicolumn{6}{|c|}{\multirow{2}{*}{\textbf{S\&P}}} &
\textbf{GMV+} & \textbf{GMV+} & \textbf{COV2+} & \textbf{COV2+} & \textbf{GMV+} & \textbf{GMV+} \\
\multicolumn{6}{|c|}{} &
\textbf{COV2} & \textbf{COV2} & \textbf{SM}   & \textbf{SM}   & \textbf{COV2} & \textbf{COV2} \\
\hline
Efficiency-A & 1.02110 & 0.96600 & 0.77240 & 1.12390 & 0.96660 & \textbf{1.08540} & \textbf{1.08540} & 1.01230 & 1.01230 & \textbf{1.08540} & \textbf{1.08540} \\
Efficiency-B & 0.95780 & 0.93970 & 0.69460 & \textbf{1.09340} & 0.88640 & 0.92060 & 0.92060 & 0.97960 & 0.97960 & 0.92060 & 0.92060 \\
Efficiency-C & 0.97400 & 1.00900 & 1.04360 & 1.01480 & 1.01060 & \textbf{1.12450} & \textbf{1.12450} & 0.96170 & 0.96170 & \textbf{1.12450} & \textbf{1.12450} \\
\hline
\multicolumn{6}{|c|}{\multirow{2}{*}{\textbf{RUSSELL}}} &
\textbf{GMV+} & \textbf{GMV+} & \textbf{GMV+} & \textbf{COV2+} & \textbf{GMV+} & \textbf{GMV+} \\
\multicolumn{6}{|c|}{} &
\textbf{QIS} & \textbf{COV2} & \textbf{QIS}   & \textbf{SM}   & \textbf{COV2} & \textbf{COV2} \\
\hline
Efficiency-A & 0.97450 & 1.02500 & 0.83650 & 1.00220 & 1.01170 & 1.01800 & \textbf{1.09560} & 1.01800 & 0.95750 & \textbf{1.09560} & \textbf{1.09560} \\
Efficiency-B & 0.97370 & 0.99760 & 0.73820 & 0.96840 & 1.00170 & \textbf{1.00310} & 0.99640 & \textbf{1.00310} & 0.95110 & 0.99640 & 0.99640 \\
Efficiency-C & 0.96160 & 0.97470 & \textbf{1.13230} & 1.01730 & 0.98310 & 1.00380 & 1.05580 & 1.00380 & 0.97150 & 1.05580 & 1.05580 \\
\hline
\multicolumn{6}{|c|}{\multirow{2}{*}{\textbf{TOPIX}}} &
\textbf{GMV+} & \textbf{GMV+} & \textbf{GMV+} & \textbf{COV2+} & \textbf{GMV+} & \textbf{GMV+} \\
\multicolumn{6}{|c|}{} &
\textbf{COVMKT} & \textbf{COV2} & \textbf{QIS}   & \textbf{SM}   & \textbf{COV2} & \textbf{COV2} \\
\hline
Efficiency-A & 0.89490 & 0.77110 & \textbf{1.19100} & 0.74380 & 1.04800 & 1.07530 & 0.99150 & 1.02810 & 0.76440 & 0.99150 & 0.99150 \\
Efficiency-B & 0.92040 & 0.82080 & 0.89660 & 0.79050 & 1.00300 & 0.98320 & 0.90810 & \textbf{1.04180} & 0.83670 & 0.90810 & 0.90810 \\
Efficiency-C & 0.92140 & 0.87940 & \textbf{1.15180} & 0.90570 & 1.01240 & 1.01060 & 1.06570 & 0.98240 & 0.80590 & 1.06570 & 1.06570 \\

\hline

\end{longtable}}
\end{center}

\begin{figure}[htbp]
    \centering

    \begin{subfigure}[b]{0.4\linewidth}
        \centering
        \includegraphics[width=\linewidth]{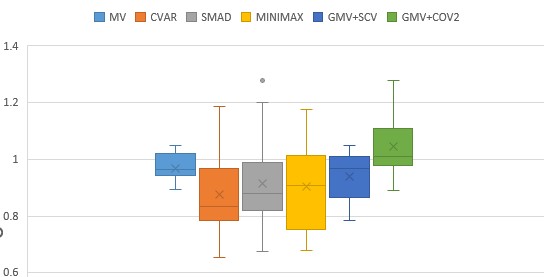}
        \caption{Group A}
        \label{fig:sub1}
    \end{subfigure}
    \hfill
    \begin{subfigure}[b]{0.4\linewidth}
        \centering
        \includegraphics[width=\linewidth]{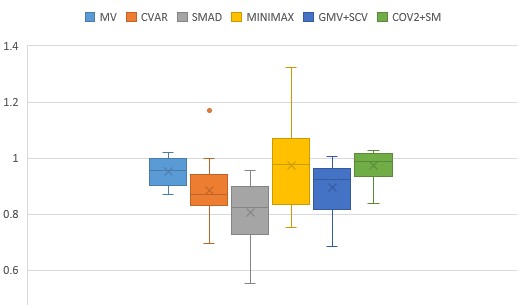}
        \caption{Group B}
        \label{fig:sub2}
    \end{subfigure}
    \hfill
    \begin{subfigure}[b]{0.4\linewidth}
        \centering
        \includegraphics[width=\linewidth]{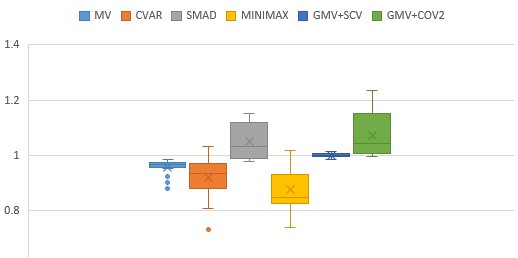}
        \caption{Group C}
        \label{fig:sub3}
    \end{subfigure}

    \caption{Efficiency scores distribution framed over different datasets and varying out-of-sample period size of the five benchmark models (MV \& GMV with sample estimates,  CVaR, MAD, MM) and the universal best-performing models corresponding to the groups A, B, and C. }
    \label{fig:combined}
\end{figure}

In closing, we present the box plot of efficiencies for the benchmark models (MV \& GMV with sample estimates,  CVaR, MAD, MM) and the universal best-performing shrinkage models across all three groups in Figure \ref{fig:combined}. The analysis includes 18 efficiency scores (6 markets × 3 out-of-sample periods) for each of the five benchmark models and the universal best-performing shrinkage model (GMV+COV2 for Groups A and B, and COV2+SM for Group C). Observations from the Figure \ref{fig:combined} are described in the summary points below. 

\begin{center}
\textbf{ Summary points}
\end{center}
\begin{itemize}
\item For group A, when all financial metrics are considered, the model GMV+COV2 performs well over the majority of the data sets, DJ, NIFTY, S\&P, and RUSSELL, while GMV+COVMKT for the TOPIX and LIS+SM for the FTSE. The GMV+COV2 model dominates all the benchmark models in terms of efficiency score as observed in Figure \ref{fig:sub1}.

\item The investigation could not identify a single shrinkage-based model for group B that performed better.  The COV2+SM model demonstrates superior performance in the DJ, NIFTY, and S\&P markets; LIS+SM excels in the FTSE; and GMV+QIS in the RUSSELL and TOPIX.   As noted in Figure \ref{fig:sub2}, COV2+SM surpasses CVaR, SMAD, and classical GMV portfolio models; however, the MM model outperforms it, while the conventional mean-variance model is comparable.

\item Analysis of group C reveals the predominance of GMV models wherein the GMV+COV2 performs the best, followed by GMV+QIS and GMV+COVMKT. The universal best-performing GMV+COV2 model is seen to outperform the benchmark PO models, MV, GMV, CVaR, and MM, as illustrated in Figure \ref{fig:sub3}. The performance of SMAD is comparable to its performance in high-dimensional datasets, primarily due to its low turnover ratio. With the dominance of GMV models for group C investors, the classical GMV model with sample covariance performs poorly in low-dimensional markets (DJ, NIFTY, FTSE) while it fits well for the S\&P 500, RUSSELL, and TOPIX.  

\item COV2 places the most effective shrinkage covariance estimators irrespective of the profile of investors and investment duration as GMV+COV2 ranks universal-best performer for the groups A and C while COV2+SM becomes universal-best for the group B. The next performance is shown by the covariance estimator QIS, as it frequently appears in top positions either with GMV or MV. 

\item It's intriguing that the sample mean SM dominates other shrinkage mean estimators as the (MV) model based on SM, i.~e, COV2+SM, becomes universal-best for group B. When the non-GMV model was the first choice for the FTSE market in group A, it was the sample mean-based MV model, which was LIS+SM. In fact, LIS+SM is the best choice for the data set FTSE over both groups, A and B. But the sample mean is never a good choice when combined with the sample covariance for any market and for any group of investors. 

\item In summary, certain shrinkage-based portfolios exhibit superior performance compared to the benchmark models across the three groups: A, B, and C. The GMV+COV2, which performs best for groups A and C, outperforms over several data sets except for FTSE (for group A \& for group C over 3-months out-of-sample period) and TOPIX (for group A) markets. 
On the other hand, the performance of shrinkage is more reliant on the underlying market structure for group B, i.~e. return-focused investors.
\end{itemize}

\section{Conclusions}\label{sec8}
This research conducts an extensive examination of Markowitz portfolio optimization using mean-variance (MV) and global mean-variance (GMV) modeling approaches, integrated with several widely-used shrinkage estimation techniques for both the mean vector and covariance matrix. The analysis employs five shrinkage estimators for the mean and eleven for the covariance matrix. Performance evaluation is conducted across six international datasets, including three high-dimensional cases where the number of assets ($p$) exceeds the number of time observations ($n$). This configuration yields sixty-six portfolio combinations for datasets with $n>p$, and forty-eight combinations when $n<p$.

Given that there are numerous portfolios to compare, we suggest utilizing a super-efficiency data envelopment analysis (DEA) model with portfolios as decision-making units (DMUs). The input and output for the DEA model are various financial metrics, such as mean return, risk measures, and reward-risk ratios. The study analyses three variations of out-of-sample window sizes: 3-months, 6-months, and 1-year, within the rolling window framework. Additionally, we categorize three groups: A, B, and C, based on the return-risk profile of investors, and perform a (DEA) ranking across the three groups. For every market and for every group,  we select the two optimal portfolios: (i) the market best performer and (ii) the universal best performer. 

The empirical analysis concludes with many interesting facts: (1) Though the choice is more dependent on the underlying market, the GMV model with the Ledoit-Wolf method of two-parameter shrinkage covariance (COV2) represents an optimal strategy for a broad spectrum of investors, though preferences shift for those prioritizing higher returns. (2) The (MV) model with COV2, together with the sample mean (SM), is more appropriate for return-seeking investors. (3) Both the models (GMV + COV2 and COV2 + SM) exhibit enhanced performance relative to conventional benchmarks such as Markowitz portfolios with sample estimates, MiniMax, CVaR, GMV, and SMAD. (4) The shrinkage covariance (COV2) often appears in top choices for all groups of investors and over several markets, establishing it as a preferred covariance estimation method,  while the sample mean (SM) shines in the mean estimation category. (5) The sample mean together with the sample covariance accounts for poor out-of-sample results, rendering it an undesirable option.

In summary, this research establishes a foundation for gaining deeper insights into how particular shrinkage models perform across varying market environments. Future research could build upon this work by investigating the effectiveness of these models in developed versus emerging markets, or in periods of stability versus periods of high volatility. 

\section*{Data Availability}
The data sources used in this study are referenced within the article.  The  intermediate findings of our empirical analysis are publicly available on GitHub:\\
https://github.com/rupendrayadav19401/Shrinkage-PO.

\section*{Acknowledgment}
The first author is grateful for the financial support received from the Council of Scientific and Industrial Research (CSIR), India (award number 09/0086(17076)/2023-EMR-I) for conducting research work. The DST-FIST grant SR/FST/MS-1/2019/45 is acknowledged for the computing facility in the department at IIT Delhi.

\appendix

\section{Benchmark Portfolio Optimization Models}
For the uniform probability vector \( P_j = \frac{1}{n}, \; j = 1, \ldots, n, \), following are the modeling structure of the three benchmark PO models considered for the comparison analysis: 

\subsection*{SMAD (Semi Mean Absolute Deviation) Model \cite{feinstein1993reformulation}}
The model introduces auxiliary variables \( d_j \in \mathbb{R} \) to represent absolute deviations from the mean portfolio return. The model is formulated as
\begin{align*}
\text{(SMAD)} \quad  \min_{\bm{x} \in X,\, d} \quad & \frac{1}{n} \sum_{j=1}^{n} d_j, \\
\text{s.t.} \quad & d_j + \bm{r}_j^\top \bm{x} \geq \bm{\bar{r}}^\top \bm{x}, \quad \forall \;j = 1, \ldots, n, \\
& d_j \geq 0, \quad \forall\; j = 1, \ldots, n.
\end{align*}
\subsection*{CVaR (Conditional Value-at-Risk) Model \cite{rockafellar2000optimization}}
Let \( \alpha \in (0,1) \) denote the confidence level. The model introduces \( \beta \in \mathbb{R} \) as the Value-at-Risk (VaR) threshold and \( u_j \in \mathbb{R} \) as auxiliary variables for losses exceeding the VaR. The CVaR model is formulated as
\begin{align*}
\text{(CVaR)} \quad \min_{\bm{x} \in X,\, \beta,\, u} \quad & \beta + \frac{1}{\alpha n} \sum_{j=1}^{n} u_j - \bm{\bar{r}^\top x}, \\
\text{s.t.} \quad & u_j \geq 0, \quad \forall\; j = 1, \ldots, n, \\
& u_j + \beta + \bm{r_j^\top x} \geq 0, \quad \forall \;j = 1, \ldots, n.
\end{align*}
\subsection*{MiniMax Model \cite{cai2000portfolio}}
It introduces an auxiliary variable \( y \in \mathbb{R} \) representing the minimum portfolio return across all scenarios. The goal is to maximize the worst-case scenario return, formulated as
\begin{align*}
\text{(MM)} \quad \min_{\bm{x} \in X,\, y} \quad & -y - \bm{\bar{r}^\top x}, \\
\text{s.t.} \quad & \bm{r_j^\top x} \geq y, \quad \forall\; j = 1, \ldots, n.
\end{align*}

\bibliographystyle{elsarticle-num}
\bibliography{sample}

@article{konno1991,
  title={Mean-absolute deviation portfolio optimization model and its applications to {T}okyo stock market},
  author={Hiroshi Konno, Hiroaki Yamazaki},
  journal={Management Science},
  volume={37},
    pages={519--531},
  year={1991},
  publisher={Elsvier}
}

@article{tran2025,
  title={Enhancing portfolio optimization in emerging markets: A cross-validation multi-target shrinkage approach},
  author={Minh Tran, Nihat M. Nguyen, 
Tuan A. Tran},
  journal={Results in Control and Optimization},
  volume={21},
    pages={100611},
  year={2025},
  publisher={Elsvier}
}

@article{jorion1986bayes,
  title={Bayes-{S}tein estimation for portfolio analysis},
  author={Jorion, Philippe},
  journal={Journal of Financial and Quantitative Analysis},
  volume={21},
  number={3},
  pages={279--292},
  year={1986},
  publisher={Cambridge University Press}
}

@inproceedings{james1961estimation,
  title={Estimation with quadratic loss},
  author={James, William and Stein, Charles and others},
  booktitle={Proceedings of the fourth Berkeley symposium on mathematical statistics and probability},
  volume={1},
  pages={361--379},
  year={1961},
  organization={University of California Press}
}

@article{bodnar2019optimal,
  title={Optimal shrinkage estimator for high-dimensional mean vector},
  author={Bodnar, Taras and Okhrin, Ostap and Parolya, Nestor},
  journal={Journal of Multivariate Analysis},
  volume={170},
  pages={63--79},
  year={2019},
  publisher={Elsevier}
}

@article{wang2014non,
  title={Non-parametric shrinkage mean estimation for quadratic loss functions with unknown covariance matrices},
  author={Wang, Cheng and Tong, Tiejun and Cao, Longbing and Miao, Baiqi},
  journal={Journal of Multivariate Analysis},
  volume={125},
  pages={222--232},
  year={2014},
  publisher={Elsevier}
}

@article{bodnar2014strong,
  title={On the strong convergence of the optimal linear shrinkage estimator for large dimensional covariance matrix},
  author={Bodnar, Taras and Gupta, Arjun K and Parolya, Nestor},
  journal={Journal of Multivariate Analysis},
  volume={132},
  pages={215--228},
  year={2014},
  publisher={Elsevier}
}

@article{ledoit2022quadratic,
  title={Quadratic shrinkage for large covariance matrices},
  author={Ledoit, Olivier and Wolf, Michael},
  journal={Bernoulli},
  volume={28},
  number={3},
  pages={1519--1547},
  year={2022},
  publisher={Bernoulli Society for Mathematical Statistics and Probability}
}

@article{ledoit2020analytical,
  title={Analytical nonlinear shrinkage of large-dimensional covariance matrices},
  author={Ledoit, Olivier and Wolf, Michael},
  journal={The Annals of Statistics},
  volume={48},
  number={5},
  pages={3043--3065},
  year={2020},
  publisher={JSTOR}
}

@article{ledoit2021shrinkage,
  title={Shrinkage estimation of large covariance matrices: Keep it simple, statistician?},
  author={Ledoit, Olivier and Wolf, Michael},
  journal={Journal of Multivariate Analysis},
  volume={186},
  pages={104796},
  year={2021},
  publisher={Elsevier}
}

@article{ledoit2003improved,
  title={Improved estimation of the covariance matrix of stock returns with an application to portfolio selection},
  author={Ledoit, Olivier and Wolf, Michael},
  journal={Journal of empirical finance},
  volume={10},
  number={5},
  pages={603--621},
  year={2003},
  publisher={Elsevier}
}

@article{ledoit2004well,
  title={A well-conditioned estimator for large-dimensional covariance matrices},
  author={Ledoit, Olivier and Wolf, Michael},
  journal={Journal of multivariate analysis},
  volume={88},
  number={2},
  pages={365--411},
  year={2004},
  publisher={Elsevier}
}

@article{chopra1993effect,
  title={The effect of errors in means, variances, and covariances on optimal portfolio choice},
  author={Chopra, Vijay K and Ziemba, William T and others},
  journal={Journal of Portfolio Management},
  volume={19},
  number={2},
  pages={6--11},
  year={1993},
  publisher={World Scientific}
}

@article{bai2011estimating,
  title={Estimating High Dimensional Covariance Matrices and its Applications.},
  author={Bai, Jushan and Shi, Shuzhong},
  journal={Annals of Economics \& Finance},
  volume={12},
  number={2},
  year={2011}
}

@article{stein1975estimation,
author="Stein, C.",
title="Estimation of a covariance matrix, Rietz Lecture",
journal="39th Annual Meeting IMS, Atlanta, GA",
year="1975",
}

@article{stein1977estimation,
  title={Lectures on the theory of estimation of many parameters},
  author={Stein, Charles},
  journal={Studies in the Statistical Theory of Estimation I},
  volume={74},
  pages={4--65},
  year={1977},
  publisher={Nauka}
}

@article{stein1986lectures,
  title={Lectures on the theory of estimation of many parameters},
  author={Stein, Charles},
  journal={Journal of Soviet Mathematics},
  volume={34},
  number={1},
  pages={1373--1403},
  year={1986},
  publisher={Springer}
}

@article{charnes1978measuring,
  title={Measuring the efficiency of decision making units},
  author={Charnes, A. and Cooper, W.W. and Rhodes, E.},
  journal={European Journal of Operational Research},
  volume={2},
  number={6},
  pages={429--444},
  year={1978},
  publisher={Elsevier}
}

@article{andersen1993procedure,
  title={A procedure for ranking efficient units in data envelopment analysis},
  author={Andersen, Per and Petersen, Niels Christian},
  journal={Management Science},
  volume={39},
  number={10},
  pages={1261--1264},
  year={1993},
  publisher={Informs}
}

@article{markowitz1952portfolio,
  title={PORTFOLIO SELECTION},
  author={Markowitz, Harry},
  journal={The Journal of Finance},
  volume={7},
  number={1},
  pages={77--91},
  year={1952},
  publisher={Wiley Online Library}
}

@article{demiguel2009optimal,
  title={Optimal versus naive diversification: How inefficient is the 1/N portfolio strategy?},
  author={DeMiguel, Victor and Garlappi, Lorenzo and Uppal, Raman},
  journal={The Review of Financial Studies},
  volume={22},
  number={5},
  pages={1915--1953},
  year={2009},
  publisher={Oxford University Press}
}

@article{jobson1981putting,
  title={Putting Markowitz theory to work},
  author={Jobson, J David and Korkie, Robert M},
  journal={The Journal of Portfolio Management},
  volume={7},
  number={4},
  pages={70--74},
  year={1981},
  publisher={With Intelligence LLC}
}

@article{Green1992,
  title={When will mean-variance efficient portfolios be well diversified?},
  author={Green, Richard C and Hollifield, Burton},
  journal={The Journal of Finance},
  volume={47},
  number={5},
  pages={1785--1809},
  year={1992},
  publisher={Wiley Online Library}
}

@book{michaud2001efficient,
  title={Efficient asset management: a practical guide to stock portfolio optimization and asset allocation},
  author={Michaud, Richard O and Ma, Tongshu},
  year={2001},
  publisher={Oxford University Press}
}

@article{scherer2007can,
  title={Can robust portfolio optimisation help to build better portfolios?},
  author={Scherer, Bernd},
  journal={Journal of Asset Management},
  volume={7},
  number={6},
  pages={374--387},
  year={2007},
  publisher={Springer}
}

@article{ledoit2012nonlinear,
  title={Nonlinear shrinkage estimation of large-dimensional covariance matrices},
  author={Ledoit, Olivier and Wolf, Michael},
  journal={The Annals of Statistics},
  volume={40},
  number={2},
  pages={1024--1060},
  year={2012}
}

@article{ledoit2015spectrum,
  title={Spectrum estimation: A unified framework for covariance matrix estimation and {PCA} in large dimensions},
  author={Ledoit, Olivier and Wolf, Michael},
  journal={Journal of Multivariate Analysis},
  volume={139},
  pages={360--384},
  year={2015},
  publisher={Elsevier}
}

@book{bai2010spectral,
  title={Spectral analysis of large dimensional random matrices},
  author={Bai, Zhidong and Silverstein, Jack William and others},
  year={2010},
  publisher={Springer}
}

@article{bodnar2019testing,
  title={Testing for independence of large dimensional vectors},
  author={Bodnar, Taras and Dette, Holger and Parolya, Nestor},
  journal={The Annals of Statistics},
  volume={47},
  number={5},
  pages={2977--3008},
  year={2019},
  publisher={JSTOR}
}

@article{feinstein1993reformulation,
  title={A Reformulation of a Mean-absolute Deviation Portfolio Optimization Model.},
  author={Feinstein, Charles D and Thapa, Mukund N},
  journal={Management Science},
  volume={39},
  number={12},
  year={1993},
pages = {1552 - 1553},
}

@article{rockafellar2000optimization,
  title={Optimization of conditional value-at-risk},
  author={Rockafellar, R Tyrrell and Uryasev, Stanislav},
  journal={Journal of Risk},
  volume={2},
  pages={21--42},
  year={2000}
}

@article{cai2000portfolio,
  title={Portfolio optimization under a minimax rule},
  author={Cai, Xiaoqiang and Teo, Kok-Lay and Yang, Xiaoqi and Zhou, Xun Yu},
  journal={Management Science},
  volume={46},
  number={7},
  pages={957--972},
  year={2000},
  publisher={INFORMS}
}

@article{best2015sensitivity,
  author    = {Best, Michael J. and Grauer, Robert R.},
  title     = {On the sensitivity of mean-variance-efficient portfolios to changes in asset means: Some analytical and computational results},
  journal   = {The Review of Financial Studies},
  volume    = {4},
  number    = {2},
  pages     = {315--342},
  year      = {2015}
}

@misc{martin2009asset,
  author = {Martin, Haugh},
  title = {Asset Allocation and Risk Management},
  howpublished = {\url{http://www.columbia.edu/~mh2078/QRM-assetallocation.pdf}},
  note = {Lecture notes: IEOR E4602: Quantitative Risk Management},
  year = {2009}
}

@article{lee2020sparse,
  author    = {Lee, Yongsuk and Kim, Min Jae and Kim, Jin Hyuk and Jang, Jae Ryoung and Kim, Woo Chang},
  title     = {Sparse and robust portfolio selection via semi-definite relaxation},
  journal   = {Journal of the Operational Research Society},
  volume    = {71},
  number    = {5},
  pages     = {687--699},
  year      = {2020}
  
}

@article{bodnar2025high,
  title={High-Dimensional portfolio selection with {HDS}h{OP} package},
  author={Bodnar, Taras and Dmytriv, Solomiia and Okhrin, Yarema and Otryakhin, Dmitry and Parolya, Nestor},
  journal={The European Journal of Finance},
  pages={1--23\;},
  year={2025},
  publisher={Taylor \& Francis}
}

\end{document}